\documentclass[12pt]{article}
\usepackage{amsmath}
\pdfoutput=1

\usepackage{graphicx,psfrag,epsf}
\usepackage{enumerate}
\usepackage{natbib}
\usepackage{url} 

\newcommand{\blind}{0}

\usepackage{etoolbox}
\newif\ifabbreviation
\pretocmd{\thebibliography}{\abbreviationfalse}{}{}
\AtBeginDocument{\abbreviationtrue}

\usepackage{xcolor}
\usepackage[titlenumbered,ruled]{algorithm2e}
\usepackage{amsfonts}
\usepackage{float}
\usepackage{subcaption}
\usepackage{datetime}
\usepackage{checkend}
\usepackage{dirtree}

\newcommand{\spewn}[1]{\text{70}}

\newcommand{\agentsn}[1]{approximately five billion}

\newdateformat{monthdate}{\monthname[\THEMONTH]}

\def\spacingset#1{\renewcommand{\baselinestretch}%
{#1}\small\normalsize} \spacingset{1}

\begin{document}

\if0\blind
{
  \title{SPEW: Synthetic Populations and Ecosystems of the World}
  \author{Shannon Gallagher, Lee Richardson, Samuel L. Ventura, and William F. Eddy\thanks{
    The authors were supported by NIGMS Cooperative Agreement U24GM110707, the MIDAS Informatics Systems Group.
 In addition, much of the computing was done on the Olympus Computing Cluster at the Pittsburgh Supercomputing Center (PSC).  We would like to thank Jeremy Espino, Sahawut Wesaratchakit, and Jay DePasse for their help in using Olympus and disseminating our ecosystems.}\hspace{.2cm}\\
    Department of Statistics, Carnegie Mellon University}
  \maketitle
} \fi

\if1\blind
{
  \bigskip
  \bigskip
  \bigskip
  \begin{center}
    {\LARGE\bf SPEW: Synthetic Populations and Ecosystems of the World}
\end{center}
  \medskip
} \fi

\begin{abstract}
Agent-based models (ABMs) simulate interactions between autonomous agents in constrained environments over time.  ABMs are often used for modeling the spread of infectious diseases.  In order to simulate disease outbreaks or other phenomena, ABMs rely on ``synthetic ecosystems,'' or information about agents and their environments that is representative of the real world.  Previous approaches for generating synthetic ecosystems have some limitations:  they are not open-source, cannot be adapted to new or updated input data sources, and do not allow for alternative methods for sampling agent characteristics and locations.  We introduce a general framework for generating Synthetic Populations and Ecosystems of the World (SPEW), implemented as an open-source \texttt{R} package.  SPEW allows researchers to choose from a variety of sampling methods for agent characteristics and locations when generating synthetic ecosystems for any geographic region.  SPEW can produce synthetic ecosystems for any agent (e.g. humans, mosquitoes, etc), provided that appropriate data is available.  We analyze the accuracy and computational efficiency of SPEW given different sampling methods for agent characteristics and locations and provide a suite of diagnostics to screen our synthetic ecosystems.  SPEW has generated  over five billion human agents across approximately 100,000 geographic regions in about 70 countries, available online.
\end{abstract}
\noindent
{\it Keywords: agent-based models,$\;$  disease modeling,$\;$  iterative proportional fitting,$\;$  parallel computing, $\;$ \texttt{R},$\;$  synthetic data}
\vfill

\spacingset{1} 


\section{Introduction to Synthetic Ecosystems}
A synthetic ecosystem is a digital representation of a population of agents  (e.g. people, livestock, cars, mosquitoes, etc.)  and their surrounding environment, typically constrained within a fixed geographic region \citep{epstein1996}.   Synthetic ecosystems are used for many purposes. For example, ABMs are used to model the spread of infectious diseases and other hypotheses (described in Section \ref{sec::intro:abm}) and require synthetic ecosystems as input \citep{Beckman1996, eubank2004nature, fred, marathe2015, porco1}. 

A synthetic ecosystem may be represented in a tabular manner as demonstrated below in Table \ref{tab::agents}.  For instance, Agent 1 is a 72 year old statistician whose environment consists of a food stand and a bathroom.  The capacity and geographic information of these environmental components are also shown.  For the purposes of this paper, an environmental component is a locus of  activity and interaction of agents in their ecosystem (e.g. schools, workplaces, hospitals, school buses, etc.).

\begin{table}[ht]
  \centering 
  \caption{Tabular representation of human agents and their environmental components in a synthetic ecosystem.}\label{tab::agents}
  \resizebox{\textwidth}{!}{

\begin{tabular}{|r|r|r|r|r|r|r|r|}
\hline 
ID &Country & Year & Occupation & Age & Gender  & Environment\\
\hline 
P1 & U.S. & 2010 & Statistician & 72 & M & E2,E3\\ 
P2 & U.S. & 2010 & Data Scientist & 54 & F  & E1,E2,E3\\ 
P3 &  U.S. & 2010 & Fisherman & 56 & M & E3\\
  \hline
\end{tabular}

\begin{tabular}{|r|r|r|r|r|}
\hline 
ID & Environment & Capacity & Latitude & Longitude \\
\hline 
E1 &Roller Coaster & 100 & 41.9276 & 77.9959 \\ 
E2&  Food Stand & 15 & 40.3782 & 79.1827 \\ 
E3& Bathroom & 50 & 41.9283 & 79.8761 \\ 
\hline	
\end{tabular}
}

\end{table}

\subsection{Agent-Based Models}\label{sec::intro:abm}
Agent-based models (ABMs) are used to simulate autonomous agents and their interactions within a constrained environment over time, and are described as a ``generative'' mode of science.  ABMs have gained significance in the past two decades, in part due to improvements in computing \citep{epstein1999agent}.  In particular, ABMs are useful when there is little information about a newly developing disease outbreak such as Ebola \citep{chan2014}, Dengue \citep{eurosurv2014}, and Zika \citep{zika2016}. By using ABMs, researchers hope to answer a variety of scientific questions including those from  epidemiology \citep{fred, porco1, epstein2011} and more.  Protocols for ABMs in ecology are given by \cite{Grimm2006}.  For a summary of ABMs in sociological instances, see the chapter by Helbing in  \cite{bonabeau2002agent}. 

Agents in an ABM represent individuals or objects with varying and often numerous characteristics.  For instance, in Table \ref{tab::agents}, each agent has a gender, age, and occupation along with the environmental components in which they may interact with one another. Synthetic ecosystems act as input to ABMs, providing the agents and their characteristics (along with the environmental components and their characteristics) to the model.  Ultimately, we want ABMs to adequately reflect reality  and provide insight that is useful for decision making \citep{bonabeau2002agent}. As such, there is a demand for synthetic ecosystems that contain agents and environments that adequately reflect reality. In the next section, we discuss some prior attempts at creating representative synthetic ecosystems.

\subsection{Previous Approaches for Synthetic Ecosystems}

One of the first sophisticated synthetic ecosystems for use in ABMs originated from the Transportation Analysis Simulation System (TRANSIMS) by \cite{Beckman1996},  which simulated traffic patterns of synthetic individuals.  There, the synthetic people were generated with the help of the Iterative Proportional Fitting (IPF)  algorithm from \cite{deming} to match certain population characteristics of synthetic individuals to U.S. Census data, which we discuss later in Section \ref{sec::sampling_methods}.  The idea behind  the synthetic ecosystem of \cite{Beckman1996} can also be seen in EpiSims \citep{eubank2004nature}, whose more recent synthetic ecosystems are specifically tailored for modeling the spread of Ebola and other diseases in Africa \citep{marathe2015}.

\begin{figure}
  \centering

\includegraphics[width=.9\textwidth]{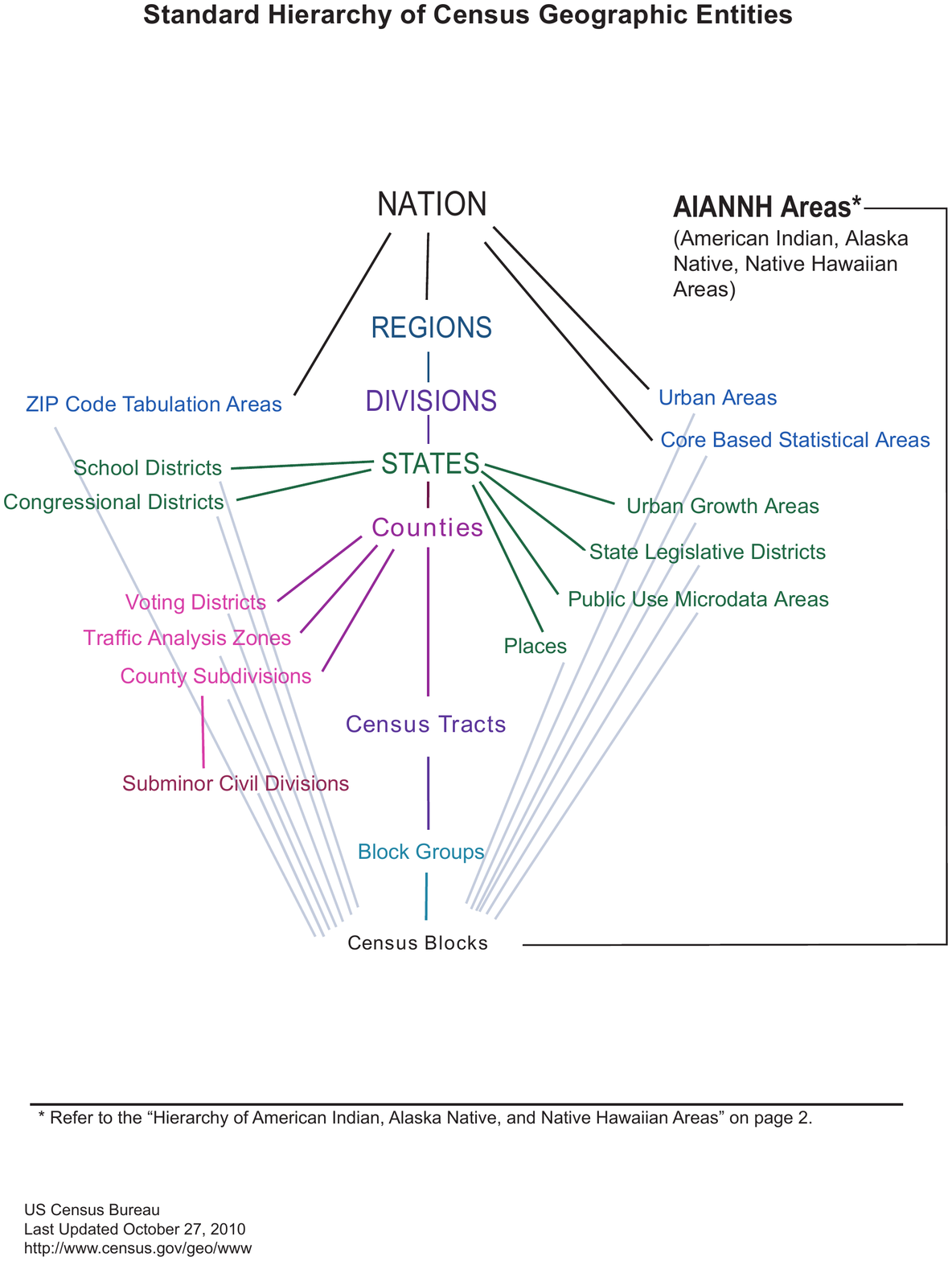}
  \caption{This figure is from the \cite{censusdiag}  and gives the geographical relationship of various U.S. regions.  The U.S. is divided into administrative units in a hierarchical fashion based on political, geographic, or population-based boundaries.  For example, a state may be partitioned into counties, which may be partitioned into census tracts.  A state may also be partitioned into Public Use Microdata Areas (PUMAs), which are the population-based equivalent of a county.  Synthetic ecosystems for census tracts can be aggregated to the county or PUMA-level.  Synthetic ecosystems at the county or PUMA-level can be aggregated to the state-level. }\label{fig::census_hier}
\end{figure}

The same idea is also seen in the \cite{Wheaton09} synthetic ecosystem.  These synthetic ecosystems were created as part of the Models of Infectious Disease Agent Study (MIDAS) network \citep{Wheaton09, wheaton2014quickstart}, and they consist of synthetic households, corresponding synthetic people, school and workplace assignments, and a separate file for group quarters (e.g. dormitories, prisons, etc.).  The populations are available for every county in the United States with data available down to the U.S. Census block group-level (see Figure \ref{fig::census_hier} pertaining to the U.S. Census geographic hierarchy for more details). The synthetic households are sampled from  American Community Survey (ACS) Public Use Microdata Samples (PUMS) data, which are a  sample of de-identified household records with a variety of characteristics, including number of persons per household, head of household age, head of household race, and household income.   Synthetic households are sampled from the PUMS, weighted by information on individual characteristics (those mentioned above) from the ACS Summary Files (SF). Locations are assigned to households through a weighted random sampling from the population density of the U.S.. Schools and workplaces are then assigned to individuals based on an adapted ``gravity model,'' which assigns agents to an environmental component based on their distance to this component and the capacity of the component.  \cite{Wheaton09} also produced interactive maps of their synthetic ecosystems.  Their populations have been used in ABMs such as FRED,  which studies the spread of diseases, especially for influenza and measles \citep{fred, porco1}.

The Wheaton synthetic ecosystem has been successfully used in ABMs,  but it has limitations.  For instance, the code is not open-source and thus is not replicable for researchers interested in generating synthetic ecosystems for other geographic regions.  Another limitation is that these synthetic ecosystems are only created for the United States, for which the input data sources are compatible and  easily converted into a standard format, or ``harmonized.''  \cite{Wheaton09} have not created extensions for other countries and have synthesized for a single year only, 2010, and so may not be representative of the U.S. population for other years.  This is particularly important because as a population evolves throughout time, researchers will be dependent on the authors creating new or updated synthetic ecosystems, with no ability to generate them on their own.  Moreover, the region size,  sampling methods, and the population characteristics of importance are all fixed.  Finally, publicly accessible diagnostic checks of the synthetic ecosystems are limited.

\subsection{Our Contribution:  SPEW}
In order to address the limitations of the previous work discussed above, we developed an open-source \texttt{R} package  called  ``SPEW: Synthetic Populations and Ecosystems of the World'' (SPEW; \texttt{R} package \texttt{spew}) for generating synthetic ecosystems.  Figure \ref{fig::spew_high_level} is a sketch of SPEW at a high level. Much of this process involves locating and harmonizing raw input data sources in order to allow the main process of SPEW to produce a synthetic ecosystem. The resulting synthetic ecosystems include synthetic agents (e.g. people) assigned to households and may contain various environmental components specific to the region of the ecosystem (e.g. schools, workplaces, disease vector populations, etc). 

\vspace{2mm}
\begin{figure}
  \centering

    \label{fig::spew_high_level}
    \includegraphics[width = 6in]{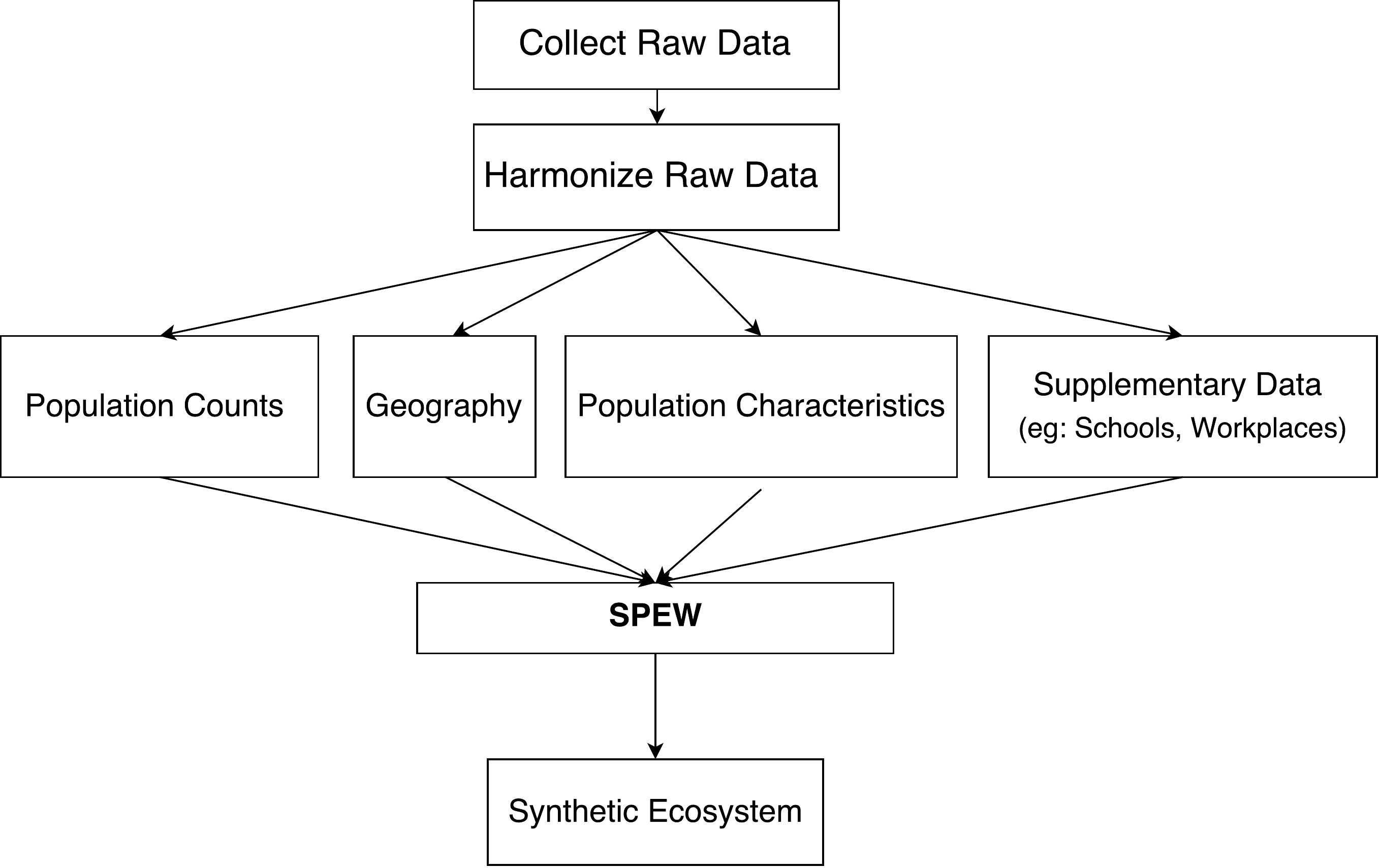}
        \caption{High level sketch of the SPEW framework.  SPEW harmonizes raw data input consisting of population counts, geographies,  population characteristics, and supplementary data.  The harmonized data is used to generate a standardized synthetic ecosystem.}   \label{fig::spew_high_level}
\end{figure}

The primary benefits of our package are three-fold:  flexible data input, enhanced user control in both data input and sampling methodology, and automatically generated summary reports for each region. For example, SPEW is not dependent on using data from a particular source or time frame; future users can use SPEW to generate synthetic ecosystems when new data becomes available (e.g. updated population counts or characteristics), and users with proprietary data (e.g. government agencies with confidential data) can incorporate this into their ecosystems with SPEW. Additionally, SPEW allows users to choose their approaches for sampling agents and locations, as described in Section \ref{sec::methods}. Finally, we automatically generate diagnostic reports that include geospatial visualizations, graphical and tabular summaries of the ecosystem for a particular region, and statistical tests to ensure that the characteristics of the ecosystem's agents are represenative of the actual ecosystem.

The rest of this paper is organized as follows. Section \ref{sec::data} distinguishes between essential and supplementary input data for synthetic ecosystems, and we discuss the challenges of harmonizing input data from many sources.  In Section \ref{sec::methods}, we overview the computational, statistical, and graphical aspects of SPEW.  In Section \ref{sec::output}, we describe the output data and overview our existing synthetic ecosystems.  In the remainder of Section \ref{sec::results}, we discuss the results of the methodology described in Section \ref{sec::methods}. Finally, we discuss the implications and potential extensions of our work in Section \ref{sec::discussion}. Appendix \ref{sec::datalist} contains supplementary material, and Appendix \ref{app::ipf} contains mathematical details about our implemented sampling methods.

\section{Input Data for Generating Synthetic Ecosystems} 
\label{sec::data}
SPEW incorporates and harmonizes input data from a variety of different sources in order to generate a  synthetic ecosystem for a given region.  In particular, SPEW relies on three essential input data types: 1) population counts; 2) geography; and 3) population characteristics.  In addition to these essential input data types, supplementary data sources may  be incorporated.        

One of the most difficult challenges we face is finding, collecting, and formatting the input data together in a systematic way.  There are two crucial steps to assembling the necessary input data for generating our synthetic ecosystems: collection and harmonization.  Because much of these processes are automated, inclusion of new and future data sources is streamlined.  As the population of the world evolves,  our input data will change (e.g. geographic boundaries, population counts, etc.), but SPEW has the ability to incorporate these new data sources to generate updated synthetic ecosystems without repeating already-completed tasks.

\subsection{Population Counts}
In order to generate a synthetic ecosystem for any region, we need to know the number of agents in that region (population counts).    Population counts are typically the easiest piece of input data to obtain.  For example, the U.S. Census Bureau  provides population counts for the U.S. for multiple administrative levels \citep{2010sf}, including for census tracts, which consist of roughly 4,000 individuals on average.  The U.S. Census Bureau also provides summary data at the tract-level and at the lower block group-level (see Figure \ref{fig::census_hier}).  We use the tract-level population counts (which are based on the decennial Census and the yearly ACS) because they typically have a smaller estimated variance than the corresponding cumulated block group counts and because some counts are suppressed at the block group-level due to privacy concerns.  

Government agencies from other countries sometimes provide population counts similar to those of the United States, but many do not.  Fortunately, \cite{geohive} provides a compendium of population counts for nearly all countries in the world.  The GeoHive population counts are typically available at multiple administrative levels (e.g. states and counties in the U.S.).  However, for many countries GeoHive only provides total population counts of individuals as opposed to both the individual and household counts provided in the United States.  Another  potential issue is that the population size of the same administrative level regions varies substantially across different countries.  Finally, as the population within a region evolves over time, an ideal system for generating synthetic ecosystems will incorporate updated population estimates.  

\subsection{Geography}
\label{sec::data_geog}
In order to generate a synthetic ecosystem for any region, we need to know the geographic boundaries of the region (geography).  These geographic boundaries often come in the form of a shapefile.  A shapefile is a machine-readable representation of the geographic boundaries of a region, including polygons, lines, points, map projections, and other data about each region (e.g. place names or area of region) \citep{esri1998shapefile}.

\begin{figure}
  \centering
  \includegraphics[width = 1.0\textwidth]{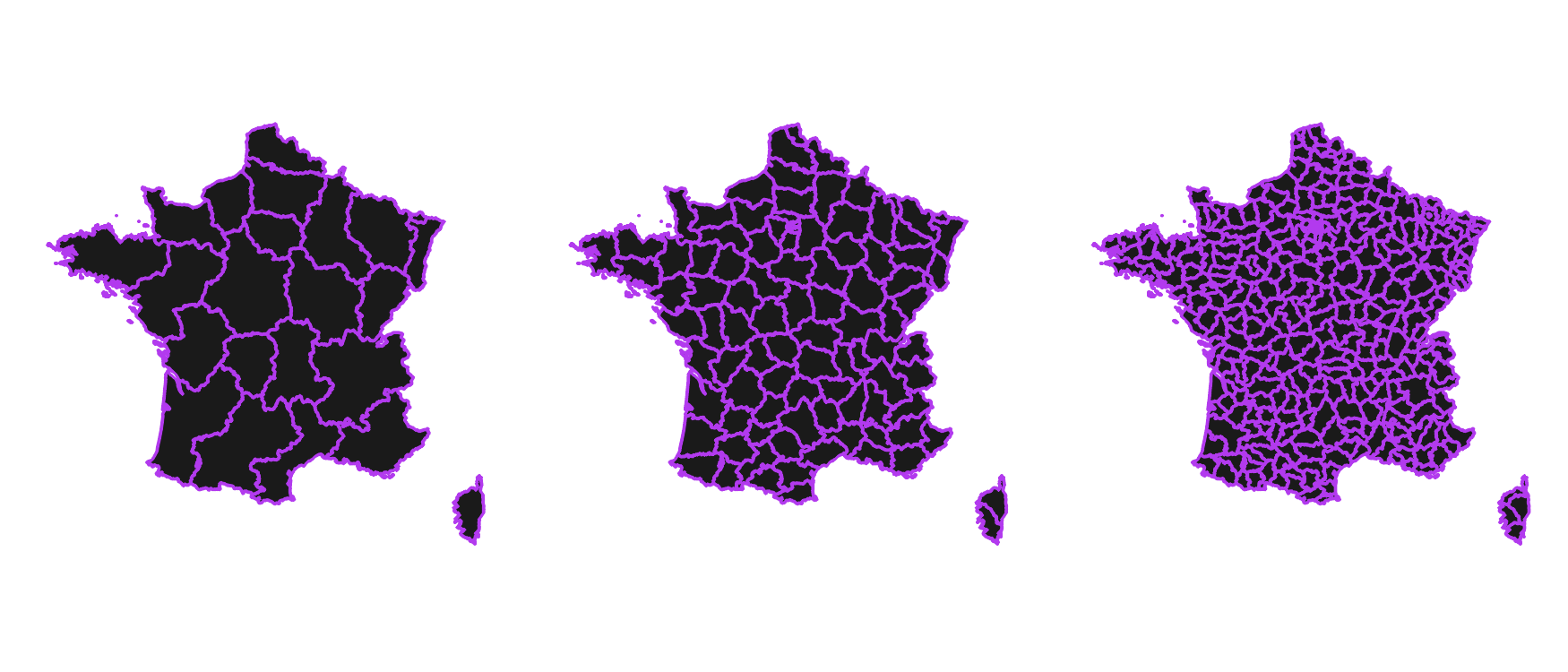}
    \caption{France partitioned at different administrative levels.  Pictured below are the Level 1, 2, and 3 administrative regions, respectively.}\label{fig::partition}
  \end{figure}

Shapefiles originate from many different sources and are typically available at multiple administrative levels for each country.  For instance, Figure \ref{fig::partition} displays three different administrative levels of France.  These particular shapefiles are from the Integrated Public Use Microdata Series, International \citep{ipumsi}, which provides shapefiles for certain countries of the world.  The U.S. Census Bureau provides shapefiles through the Topologically Integrated Geographic Encoding and Referencing (TIGER) system down to the tract-level \citep{ustiger2010}.  Additionally, the U.S. Census Bureau and TIGER provide road locations in addition to administrative boundaries for every region in the United States.

Geographic boundaries of many countries or regions are dynamic over time.  For example, the (theoretical) United States shapefiles in 1776 would include 13 regions and few sub-regions.  Today, there are 50 states and over 88,000 census tracts.  In the future, some tracts will be split or combined depending on population trends throughout time.  An ideal system for generating synthetic ecosystems will adapt to dynamic geographic boundaries.  Using shapefiles and spatial sampling, SPEW assigns each agent a latitude and longitude within their geographic region.  We discuss the different approaches to sampling in Section \ref{sec::spatial_sampling}.

\subsection{Population Characteristics}
In order to generate a useful synthetic ecosystem for any region, we need to know some characteristics of the individuals in that region (e.g. age, race, gender, and income).  For example, age is considered an important characteristic in disease modeling for humans, since the likelihood of fatality after contracting a disease depends on a person's age \citep{daley2001epidemic}.  As discussed in Section \ref{sec::sampling_methods}, there are different methods to generate the characteristics of synthetic individuals.  Each of these methods requires different types of data, discussed below.

\subsubsection{Population Microdata}
Microdata contains a single record for each person in the sample that includes individual characteristics (e.g. age, race, gender, income) and other important characteristics (e.g. household ID and household structure).  Microdata, typically in the form of PUMS, contain a small sample of the population in a region (typically about 1\% in the U.S.).  Ideally, microdata are available for each region that we wish to generate in our synthetic ecosystem.

For the U.S., we obtain microdata from the ACS at the PUMA-level \citep{5yearuspums2010}.  PUMAs are the union of tracts (see Figure \ref{fig::census_hier}) and typically have more than 100,000 individuals.  For almost all other countries, microdata comes from \cite{ipumsi}.  It would be hard to overstate the importance of IPUMS-I in generating SPEW synthetic ecosystems. IPUMS-I contains harmonized microdata for dozens of countries, along with the corresponding geographies.  Finally, our Canada synthetic ecosystem uses census information from \cite{statcan}, which provides microdata separately for major metropolitan (e.g. Montreal) and rural areas. 

When microdata is available, we can obtain both marginal and joint distributions of population characteristics for use in sampling.  We may also directly sample individuals from the microdata to preserve any joint dependencies between the characteristics of a region's population.

When microdata is not available, we still need to know the characteristics of a region's population.  One solution to this may be to use publicly available tables providing the joint distribution(s) of two or more population characteristics in a given region so that we can sample individuals with appropriate characteristics for that region.  However, in the U.S., few such tables are published at the tract-level, and in other countries, this information is rarely available at all.  Another solution is to use the marginal distributions of individual characteristics and the statistical techniques discussed in Section \ref{sec::sampling_methods} to generate individuals with appropriate characteristics.  The U.S. Census Bureau provides marginal distributions for a wide variety of characteristics at the tract-level via the ACS.  In other countries, the availability of marginal distributions at different administrative levels varies substantially.  If available, marginal distributions may be a useful alternative to using microdata in cases when the marginal distributions are available at a finer administrative level than the microdata.

\subsection{Supplementary Data}
The three data sources discussed above are essential to generating a synthetic \textit{population}.  In order to generate a more detailed synthetic \textit{ecosystem}, we need to assign agents to supplementary environmental components (e.g. schools, workplaces, hospitals, etc.) where activities take place.  Additionally, generating and incorporating synthetic populations of disease vectors may improve researchers' ability to model the spread of vector-borne diseases.

In the United States, we have schools (locations, capacities) available from the National Center for Education Statistics \citep{elsi2015} and workplaces (locations, capacities) from the Economic and Social Research Institute \citep{esriBusiness}, which we incorporate into our U.S. synthetic ecosystems.  To date, we have been unable to obtain a complete set of schools and workplaces for any non-U.S. countries.  As discussed in Section \ref{sec::place_assignments}, we can incorporate new data sources into our synthetic ecosystems as they become available.

\section{Methods for Generating Synthetic Ecosystems}
\label{sec::methods}
After harmonizing the various input data sources, we are able to generate synthetic ecosystems.  At each step of this process, there are several statistical and computational choices that can be made.  Our framework guides the generation of synthetic ecosystems in the presence of these choices.  In the following sections, we describe the SPEW framework and the corresponding methodological choices that can be made. 

\subsection{SPEW: A Framework for Generating Synthetic Ecosystems}

The main insight of SPEW is that once input data are harmonized, processing input data to an output synthetic ecosystem is similar across regions.  A sketch of how we have abstracted this process is contained in Algorithm \ref{alg::spew}.

\begin{algorithm}
\caption{Process for Generating Synthetic Ecosystems with SPEW } 
\label{alg::spew}
\SetKwInOut{Input}{input}\SetKwInOut{Output}{output}
\SetAlgoLined
\Input{Harmonized counts, geographies, population characteristics and supp. data}
 \For{Every region} {
  1. Sample population characteristics of households \\
  1. Sample population characteristics of agents \\
  2. Sample locations of agents \\
  3. Add environmental components (e.g. schools, workplaces, vectors, etc)\\
  \Output{Synthetic ecosystem for region} 
  }
\end{algorithm}

\indent After formatting the input data described above, we generate a synthetic ecosystem for each region independently (ideally, in parallel as described in Section \ref{sec::comp}).  First, we sample population characteristics of households.  Next, we sample population characteristics of agents belonging to the sampled households.  The sampling methodology is straightforward from a statistical perspective.  For example, we use methods such as simple random sampling  and iterative proportional fitting, see Section \ref{sec::sampling_methods}.  Next, we sample locations of agent locations.  We may sample agents uniformly across the region or according to some probability distribution over a subspace of the region (e.g. along roads in a U.S. state), see Section \ref{sec::spatial_sampling}.  Finally, we assign agents to environmental components, if supplementary data is available.  When assigning agents to environmental components, we may use modified versions of the gravity model from \cite{Wheaton09}.  The result is a synthetic ecosystem for each region that can be combined into a single, full synthetic ecosystem for all regions contained in the input data.

The SPEW framework provides a flexible, modular approach to the generation of synthetic ecosystems.  SPEW's flexibility gives several advantages.  First, we are able to efficiently adapt to new data sources.  In addition, it is straightforward to substitute different sampling methodologies.  This allows us to choose methodology most suitable for the input data, and quickly incorporate new and improved sampling methodologies as they become available. 

\subsection{Approaches for Sampling Population Characteristics}
\label{sec::sampling_methods}

After using the population count input data to determine the number of agents to sample, we sample population characteristics. There are many ways in which population characteristics may be sampled, and we describe three approaches here:  1) Simple Random Sampling (SRS)  2) Moment Matching (MM) and 3) Iterative Proportional Fitting (IPF).  The goal of these sampling methodologies is to produce the most realistic synthetic ecosystems possible, subject to various constraints (e.g. data availability). Below, we detail the methodology and discuss the advantages and disadvantages of each approach.  

\subsubsection{Simple Random Sampling (SRS)}
For our purposes, SRS is equivalent to uniformly sampling with replacement from the microdata.  The advantages of this approach are that it is easy to implement and interpret. The marginal and joint distributions of the sampled characteristics should match those of the microdata.  Another advantage is that SRS only requires microdata, available for many countries.  One potential disadvantage is that this approach assumes that the marginal and joint distributions in the microdata are representative of the sampled region, which may not be true.  For example, in the U.S., we sample agents at the tract-level, using microdata from the PUMA-level, a superset of the tracts (as shown in Figure \ref{fig::census_hier}).  If the marginal and joint distributions of the PUMA population characteristics do not match those of the tracts, the sampled population characteristics will not be representative of the true population (see Section \ref{sec::comp_agents}).

\subsubsection{Moment Matching (MM)}
As mentioned in Section \ref{sec::data}, we do not have microdata for every region.  In cases such as these, researchers may use microdata from similar regions to generate a synthetic ecosystem.  MM is an approach to ensure that the generated synthetic ecosystem is representative of the true population in such cases.  Assuming that the first moment of a population characteristic is available (e.g. mean age of the region's population), MM weights the microdata records of a similar region so that after sampling is completed, the corresponding moment in the resulting weighted random sample matches the known moment.  More details may be found in Appendix \ref{app::ipf}.

We find MM to be useful in cases such as The Gambia, where we only had microdata from a nearby country.  Household size was an important characteristic in this application, where Ebola was the disease of interest.  However, the average household size in the neighboring microdata was significantly different than that of The Gambia, resulting in the wrong population size (of individuals) in The Gambia.  We used MM to correct the population size by matching on the household average of the The Gambia.  More data is required for MM than in SRS, but only the first moment of a characteristic is required. This is in contrast to IPF below, which requires knowing the entire distribution of a population characteristic.  Currently, we only use MM for continuous or ordinal population characteristics. 

\subsubsection{Iterative Proportional Fitting (IPF)}
First introduced by \cite{deming} and with statistical properties given by \cite{fienberg1970}, IPF estimates the joint distribution of population characteristics given their marginal distributions and, ideally, an estimate of the covariance structure among these characteristics.  Specifically, IPF estimates individual cell values of a contingency table with known marginal totals under the assumption of non-zero cells.  Moreover, a seed table is used to initiate the IPF algorithm, which allows us to incorporate prior information about the joint distribution of the characteristics.  Given the estimated contingency table, \cite{Beckman1996} designed a scheme to sample population characteristics.  Their two-step method works as follows:  1) estimate the joint distribution of population characteristics with IPF; and 2) sample agents from microdata using the estimated joint distribution probabilities as weights.  More details may be found in Appendix \ref{app::ipf}.

The primary advantage of using IPF is that the marginal distributions of the generated population characteristics will be accurate at finer geographic levels.  For example, when microdata is available at the PUMA-level but marginal distributions are available at the tract-level, IPF will more closely match the tract-level marginal distributions than SRS.  We demonstrate this in Section \ref{sec::comp_agents}.  An additional advantage of IPF is that the user may choose which population characteristics to emphasize, ensuring that these marginal distributions are most accurately represented in the sampled agents. 

The main disadvantage of IPF is that it requires marginal distributions in addition to microdata.  At the moment, marginal distributions of population characteristics at a low administrative level are unavailable for most countries.  For example, while the U.S. Census Bureau provides this information at the tract-level, to the best of our knowledge, this information is not publicly available for countries such as Russia and China.  Additionally, if microdata is unavailable, IPF assumes independence among the population characteristics.  Finally, since the additional step of estimating the joint distribution is necessary, IPF is more computationally complex than SRS. 

\subsection{Spatial Sampling of Agents}
\label{sec::spatial_sampling}
After determining the characteristics of our agents, we next assign each agent to a location within the region.  We currently have three practical approaches to doing this, each of which fits into a simple, general framework:  given a probability distribution over the geography of a region, we sample agents according to that distribution.  The three approaches differ only in how they specify this probability distribution, namely:  1) uniformly across a region (polygons); 2) uniformly along roads (lines); and 3) according to a given spatial distribution of agent locations (weighted polygons or lines).  Each of these methods is subject to the data available for a region.

\subsubsection{Sampling Uniformly Across a Region}

When we do not know the detailed spatial distribution of agents in a region, one simple approach is to sample agent locations uniformly across a region as the boundaries of a region are stored in a shapefile. 

\begin{figure}
  \centering
  \includegraphics[width=6in]{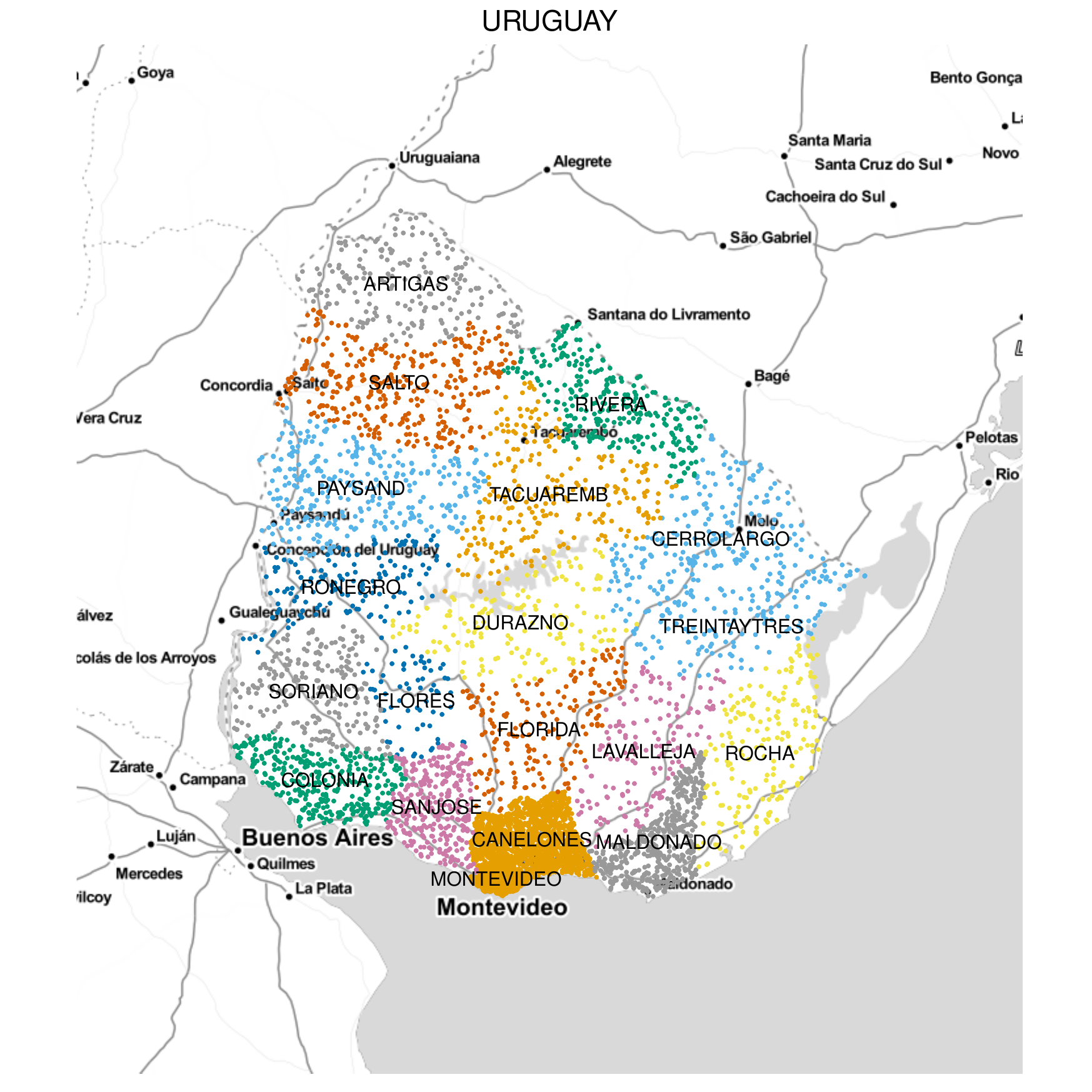}
    \caption{Map of a sub-sample of 10,000 households from our synthetic Uruguay population of about 3.3 million individuals whose locations are uniformly distributed within each sub-region.  Although population density is maintained at a macro-level, the population density within each region is uniform and likely does not reflect the true micro-level population density.}\label{fig::uruguay_map}
\end{figure}

Of course, while assuming a uniform distribution of agents in a region is unrealistic in real-world applications (as we show in Section \ref{sec::roads_methods}), this approach does maintain the macro-level population density characteristics of broader regions.  For example, Figure \ref{fig::uruguay_map} shows the results of applying the uniform sampling scheme when generating a synthetic ecosystem with SPEW for the country of Uruguay, which has 19  sub-regions.  As we see in the figure, densely populated regions such as the capital, Montevideo (bottom of map), are also densely populated in our synthetic ecosystem of Uruguay.  Conversely, agents in rural areas such as Artigas (near top of map) are sparsely distributed.  While the micro-level population density of these sub-regions is likely inaccurate, the macro-level population density of Uruguay is maintained.  We note that this approach is the only choice for spatial sampling of agents when additional geographic information is unavailable.  A disadvantage of this method is that this method disregards the surrounding topography:  we may have assigned agents to reside in lakes, deserts, or other uninhabited areas.

{}
\subsubsection{Sampling Uniformly Along Roads}\label{sec::roads_methods}
When we have information on the spatial distribution within the region, we may construct a more realistic probability model of the spatial distribution of agents in the region.  For example, if we know where roads are located within a region, we may construct a probability distribution for agent locations on the roads.  As discussed in Section \ref{sec::data_geog}, we know where roads are located in the United States, unlike other countries. 

The advantages of this approach are obvious:  people typically live near roads.  By using road-based sampling, we can assign agents to more realistic locations.  We compare road-based sampling to uniform sampling in Section \ref{sec::results_spatial}.  An additional advantage is that road-based sampling ensures that both the micro- and macro-level spatial distributions of agents are more closely representative of their true spatial distributions.  Also, this approach is easily adaptable, such as excluding interstate highways from the set of sampled roads.

A disadvantage of this approach is that it requires data that is unavailable for most countries or regions. Additionally, even when this data is available, we still may sample agents to live on bridges, in tunnels, or in other non-residential areas.  Finally, some human populations are completely removed from roads and thus would not be correctly represented in our synthetic ecosystem.

\subsubsection{Sampling According to a Known Spatial Distribution}
An option closely related to road-based sampling is sampling according to another known spatial distribution of agents. For example, \cite{Wheaton09} uses the Integrated Climate and Land-Use Scenario's (ICLUS) from the Environmental Protection Agency \citep{iclus} to estimate the spatial distribution of agents in the United States.  The advantages and disadvantages of this approach are similar to those of road-based sampling.  An additional disadvantage of this approach is that these data sources are often costly, difficult to obtain, and/or proprietary. 

\subsection{Assigning Synthetic Agents to Environmental Components} 
\label{sec::place_assignments}
Besides including population characteristics in a synthetic ecosystem, it is often beneficial to include other environmental components where the agents may interact.  For the purposes of infectious disease ABMs, it seems essential to include information on schools and workplaces, since these are the primary environmental components where diseases are transmitted \citep{fred}.  Examples of other useful environmental components for infectious disease ABMs include places of worship, airports, restaurants, and hospitals.

To allow for the incorporation of these environmental components in SPEW synthetic ecosystems, we have adapted a gravity model to assign agents to environmental components when information on these components is available.  We make the assumptions that agents are more likely to go to:  1) environmental components that are physically closer to their household location; and 2) environmental components where many other agents visit.

Let $e_i$ $i = 1, 2, \dots, P$ be environmental components with locations $l_i$ and capacity $c_i$. For an agent $x_j$ with location $l_j$, the probability that agent $x_j$ is assigned to $e_i$ is
\begin{align*}
P(x_j \textnormal{ is assigned to } e_i) &\propto d(l_i, l_j)^{-1} \cdot f(c_i),
\end{align*}
where $d$ is the distance between two locations, and $f$ is a monotonic increasing function of the capacity of $e_i$. For $d$, we use the spherical law of cosines and a step function for $f$.

\begin{figure}
  \centering
  \includegraphics[width=1.\textwidth]{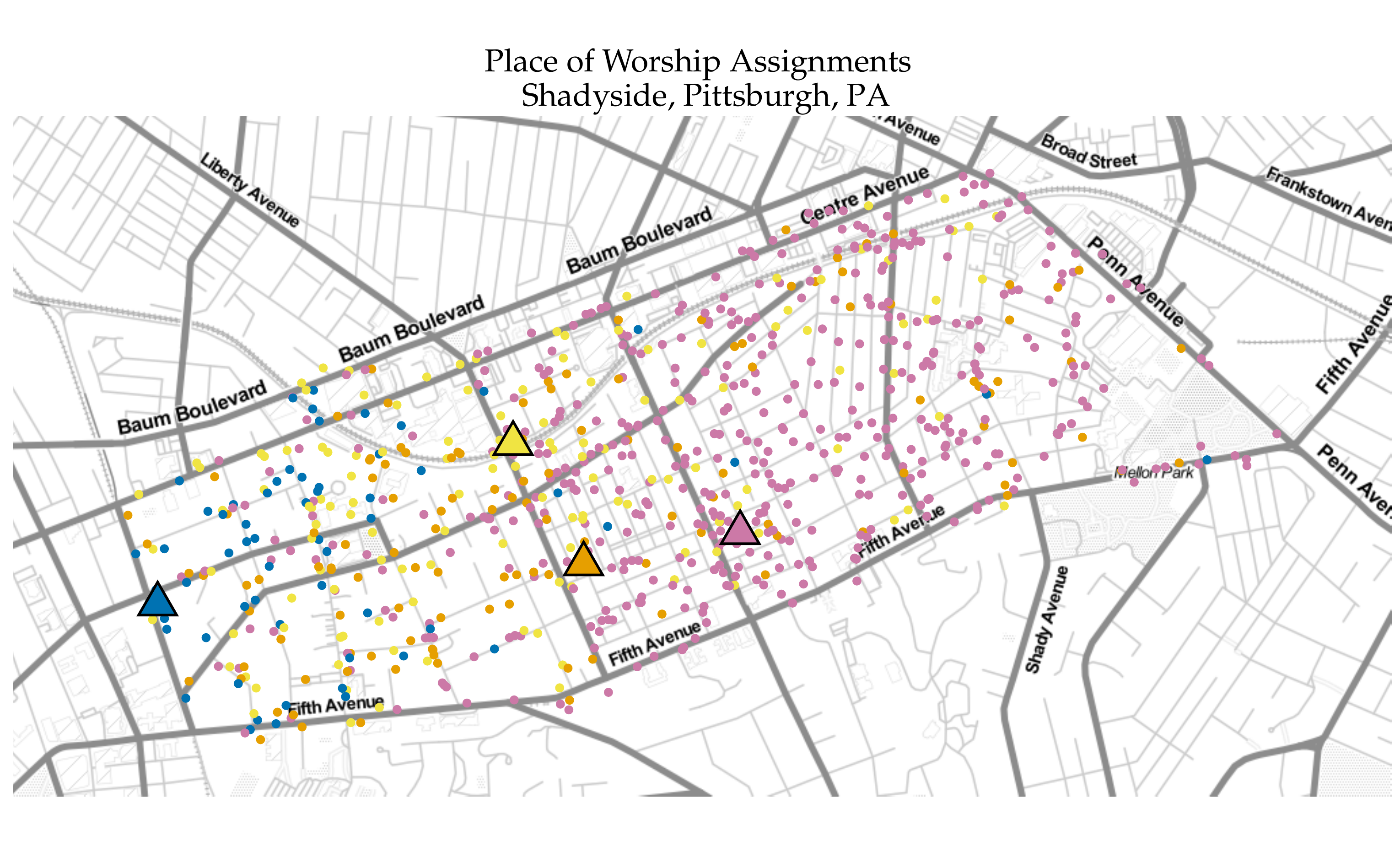}
    \caption{Place of worship assignments in Shadyside, Pittsburgh, PA.  The assignments are based both on physical distance from the place of worship ($\bigtriangleup$) and the capacity.  We assigned 800 agents ($\bullet$) of the approximately 8,000 household locations in the region to attend a place of worship indicated by one of four colors.}\label{fig::churches}
\end{figure}

Figure \ref{fig::churches} provides an illustrative example of how this method of environmental component assignment works.  Using OpenStreetMap \citep{openstreetmappaper}, we obtain the locations of places of worship in four Pittsburgh, PA census tracts.  Our approach assigns agents to attend places of worship close to their household locations.  In addition, we see that the place of worship identified by a blue triangle was assigned fewer agents than the other places of worship, while the place of worship identified by a purple triangle has the largest number of assignments.

\subsection{Automated Diagnostics:  Analysis and Visualization}
After our synthetic ecosystems are generated by SPEW, we verify their accuracy through both visualizations  and statistical analysis of population characteristics.  Our diagnostics are split into two parts: 1) user-friendly region reports; and 2) large-scale, automated statistical analyses.  These diagnostics are run on each generated region, typically at the country-level with the lower regions summarized within.  The diagnostics provide the user a brief summary of the region of interest as well as a more technical analysis for those interested in the fine details of SPEW output.

\begin{figure}[h]
  \centering
  \includegraphics[width=6in]{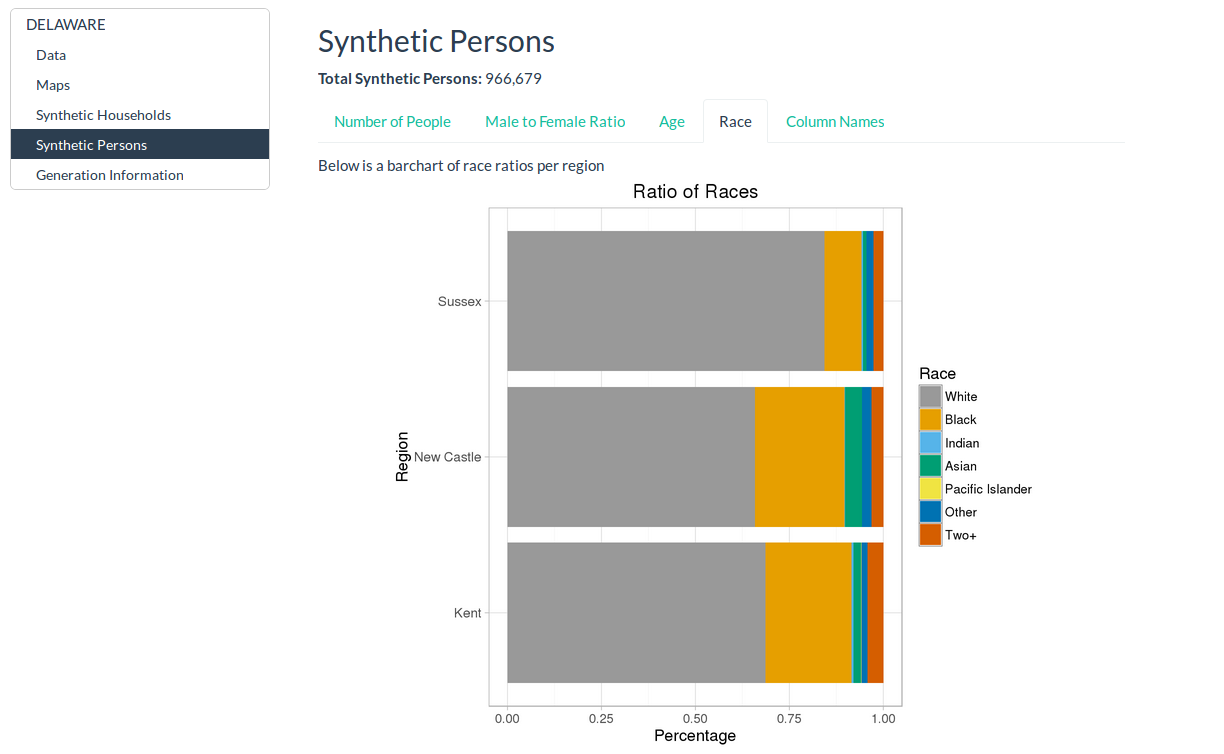}
    \caption{Screenshot of part of the automated report generated for Delaware.}\label{fig::diags_report}
\end{figure}

The region reports include helpful summaries of each region and include a map of a sample of synthetic agents, the number of agents per region, barplots of certain population characteristics, which characteristics are available in the resulting synthetic ecosystem, and generation information.  These reports (a section of which is shown in Figure \ref{fig::diags_report}) may be used as a high-level overview of the synthetic ecosystem.  
Complementary to our user-friendly reports are statistical analyses of our ecosystems to verify their realism and quality.  Our agents' population characteristics should be representative of the input data used to generate the synthetic ecosystem.  We use the Pearson goodness-of-fit $\chi^2$ statistic to test whether the distribution of categorical variables of the synthetic agents matches those of the microdata.  In the United States, we end up performing these tests on approximately 88,000 tracts for various combinations of population characteristics (e.g. race, age, gender, income).  Due to the large number of tests, we adjust for multiple comparisons using the Bonferroni correction as reported in \cite{bonferronimc}.  We summarize the results of these tests in Section \ref{sec::results}.

\subsection{Addressing Computational Challenges in SPEW}
\label{sec::comp}
Generating approximately 100,000 synthetic ecosystems is a computationally intensive task, requiring careful consideration of various issues.  In particular, we want SPEW to be:  1) efficient in speed and memory; 2) automated to reduce manual labor; and 3) organized into an \texttt{R} package for reproducibility. 

The computational efficiency of SPEW synthetic ecosystems is largely due to our access to the Olympus computing cluster, hosted at the Pittsburgh Supercomputing Center.  Olympus consists of 65 nodes, 3,872 logical cores, and over 10TB of aggregate memory (further details may be found at \cite{olympus}).  Olympus enhances the computational efficiency in two major ways.  First, it provides the memory required to compute and store our synthetic ecosystems. Second, the large number of cores makes it highly suitable for parallel computation. 

To take advantage of these cores, SPEW relies on parallelization. Specifically, SPEW partitions larger regions (e.g. countries) into sub-regions (e.g. states or tracts) and generates a synthetic ecosystem for each sub-region (see Algorithm \ref{alg::spew}).  Since generating the synthetic ecosystem for each sub-region is independent of others, each sub-region may be generated on a single core.  Determining how to partition the region/allocate the cores is very important for computational efficiency.  For example, Figure \ref{fig::partition} shows different ways of partitioning France into sub-regions.  Section \ref{sec:comp-scal-spew} analyzes the run-time of SPEW synthetic ecosystems under different conditions.

\section{Results}\label{sec::results}

\subsection{Output Data from SPEW}\label{sec::output} 
For any region, SPEW outputs at least two tables:  an agent-level file and a household-level file, released as comma-separated value (.csv) files, indexed by their region identification and SPEW version information.  The rows of the agent-level table correspond to individual agents and the columns correspond to their sampled characteristics, including a household identifier and a multitude of population characteristics including race, age, gender, income, and any other characteristics chosen by the user from those available.  The rows of the household-level file correspond to individual households while the columns correspond to household characteristics, including the household identifier, the physical location of the household (i.e. latitude and longitude), and any household characteristics supplied by the user.  These two tables are linked by the household identifier.  If environmental components are assigned to the agents, additional information about these environmental components is included in the agent-level file, and additional tables for each type of environmental component are produced. 

\begin{figure}
  \includegraphics[width=\textwidth]{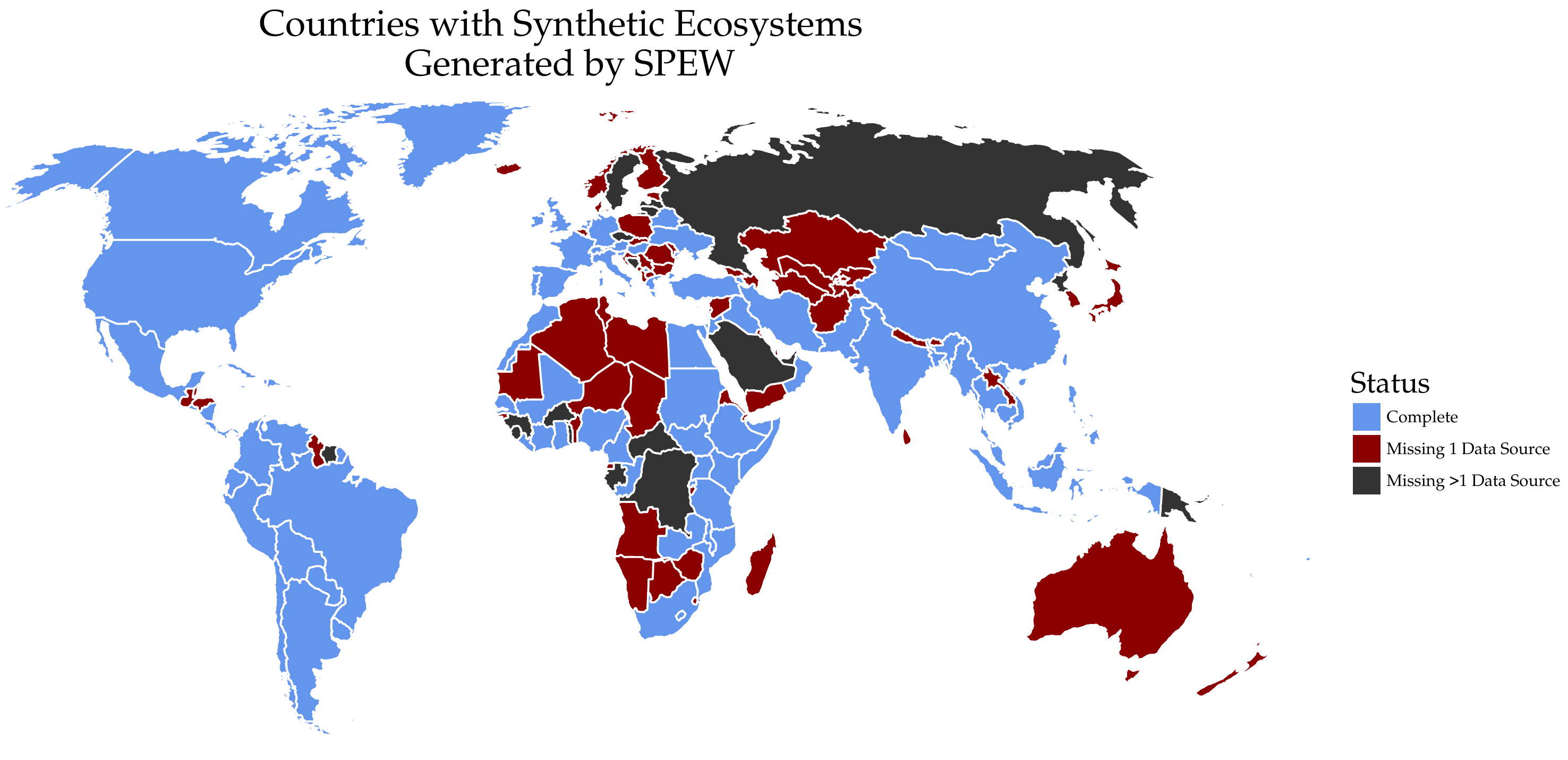}
    \caption{Map of countries with SPEW synthetic ecosystems as of \monthdate{\today} 2016.  The synthetic ecosystems of the \spewn{} countries colored in blue are completed by SPEW and available online.  The countries colored in red are missing a single essential data source, while the countries colored in gray are missing multiple essential data sources.  Once these data sources are obtained, SPEW may generate synthetic ecosystems of these countries.    The completed synthetic ecosystems are available at \small\protect\url{http://www.epimodels.org/drupal-new/?q=node/112}.}
  \label{fig::spew_world}
\end{figure}

Currently, we have produced \agentsn{} unique agents from over \spewn{} countries.  From Figure \ref{fig::spew_world}, we see that most of the Western Hemisphere is complete due to the relatively easy access of the essential input data,  but the Eastern Hemisphere, where essential data input is more difficult to obtain, is relatively incomplete.  We are missing only one essential input data source for countries colored in red (e.g. Australia).  For countries colored in black (e.g. Russia), we are missing multiple essential input data sources.  The countries colored in blue (e.g. United States) are completed and available online.

Finally, we note that SPEW is a general program for generating synthetic ecosystems for any type of agent and any type of environment or location.  As such, we do not directly synchronize our synthetic ecosystem output with existing geographic identification systems such as the Apollo Location Services \citep{wagner2013apollo}, aside from those used as input to SPEW.  That said, SPEW output may be linked to other geographic identification systems.  For example, our pre-made synthetic ecosystems are available at {\small \url{http://www.epimodels.org/drupal-new/?q=node/112}} and are searchable by region name or ISO-code via the Apollo Location Services.

\subsection{Assessing Sampling Population Characteristics Approaches}\label{sec::comp_agents}  
In Section \ref{sec::sampling_methods}, we described three methods of sampling population characteristics:  SRS, MM, and IPF.  We do not show the results of MM here as our input data includes a nominal characteristic, for which, at the moment, MM has no analogue.  Here, we assess the accuracy of SRS and IPF in terms of how well they match the input data in two ways:  1) comparing the mean absolute error and 2) using the Pearson $\chi^2$-test to determine whether the marginal distributions of our sampled population characteristics have the same distributions as those of the input data.  For instance, \cite{stevenson2010} reported how some U.S. PUMS data have ``serious problems with age and sex,'' and as a result it is useful to see whether we find these distributions to be realistic.

\subsubsection{Mean Absolute Error}
\begin{figure}
  \centering

\begin{subfigure}{.5\textwidth}
  \centering
  \includegraphics[width =\linewidth]{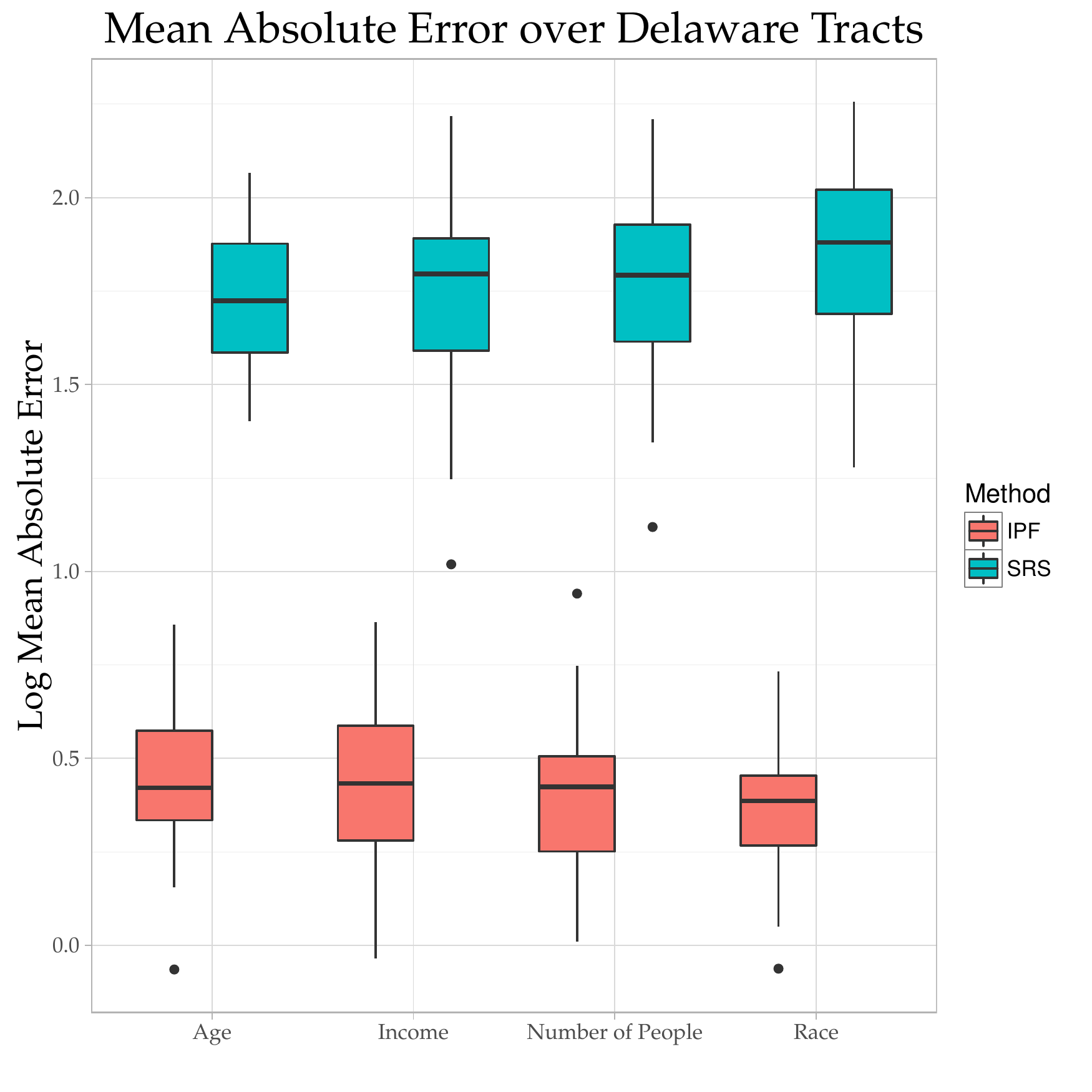}
    \caption{}  \label{fig::marginal_allsd}
\end{subfigure}%
\begin{subfigure}{.5\textwidth}
  \centering
  \includegraphics[width = \linewidth]{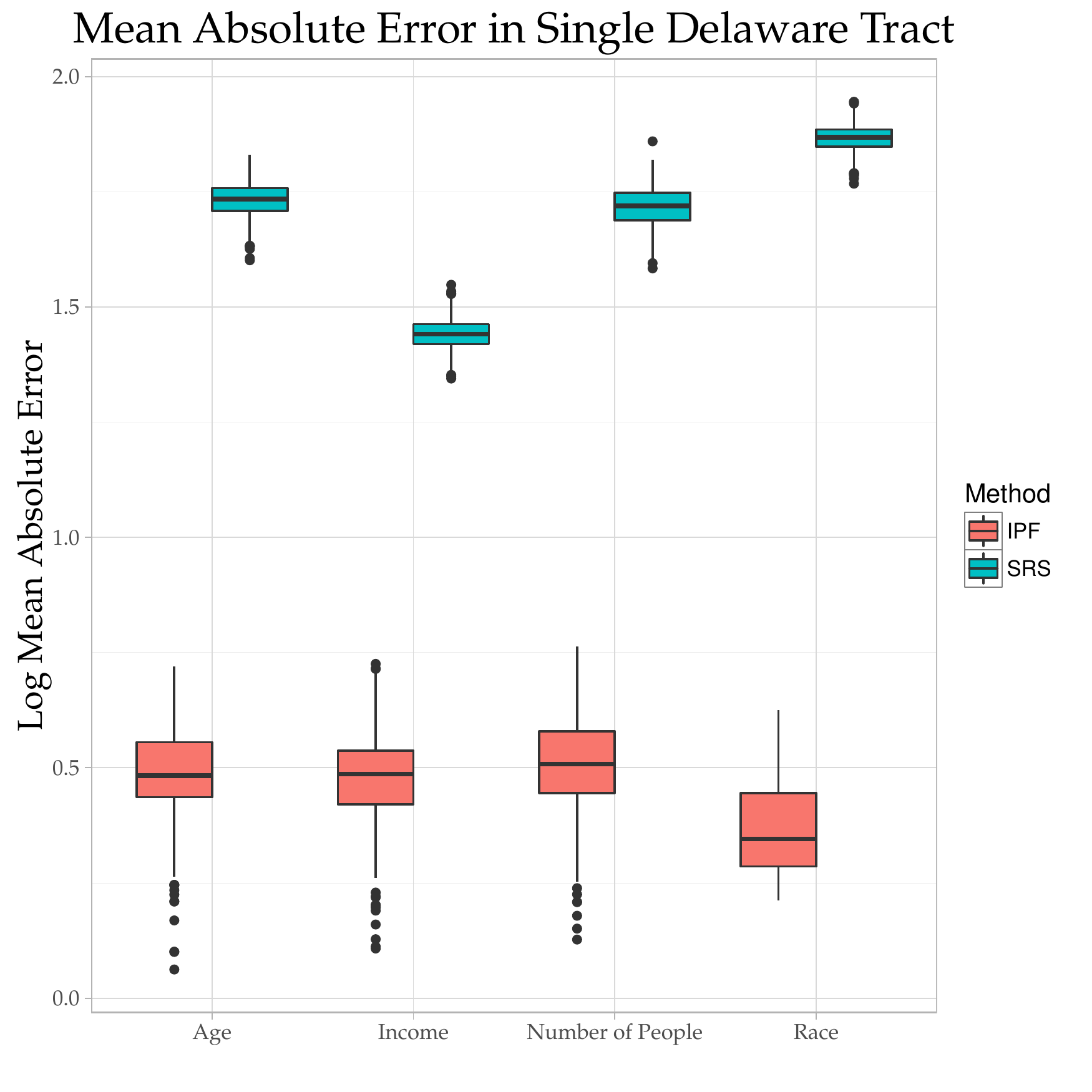}
    \caption{}  \label{fig::single_tract}
\end{subfigure}
  \caption{Left:  Log MAE between the true marginal distribution and the SPEW-generated marginal distribution across all 214 Delaware census tracts by population characteristic (age, race, household income, and household size), and sampling method (SRS or IPF).  Right:  Log MAE between true marginal distribution and SPEW-generated marginal distribution of a single Delaware tract.  SPEW was run 10,000 times in order to examine the variation of MAE within a single tract by sampling method.}\label{fig::mae}
\end{figure}

A straight forward approach to assess the accuracy of population characteristics is to calculate the mean absolute error (MAE) between the known and generated marginal distributions across all regions.  We use MAE (as opposed to mean squared error) due to its ease of interpretability.  Figure \ref{fig::marginal_allsd} shows the MAE across four known marginal distributions (age, race, income, and household size) for all tracts in Delaware.  We see that IPF has a much lower MAE than SRS.  This is because IPF explicitly matches the known marginal distribution of each tract, whereas SRS only uses data at the PUMA-level. 

Another consideration when sampling population characteristics is that the sampling methods are random.  To see this, note that both SRS and IPF assign weights to microdata records and then randomly sample according to these weights.  To understand the variability inherent in each sampling method, we regenerate the synthetic ecosystem for a single tract 10,000 times.  In each of these, IPF  samples population characteristics according to the weights it assigned to records in the microdata, while SRS uses an unweighted random sample.  Figure \ref{fig::single_tract} shows that the MAE of IPF has more variability than SRS, but is consistently smaller, with no overlap.  

\subsubsection{Comparing Marginal Distributions}
\begin{figure}
  \centering
  \includegraphics[width = .9\linewidth]{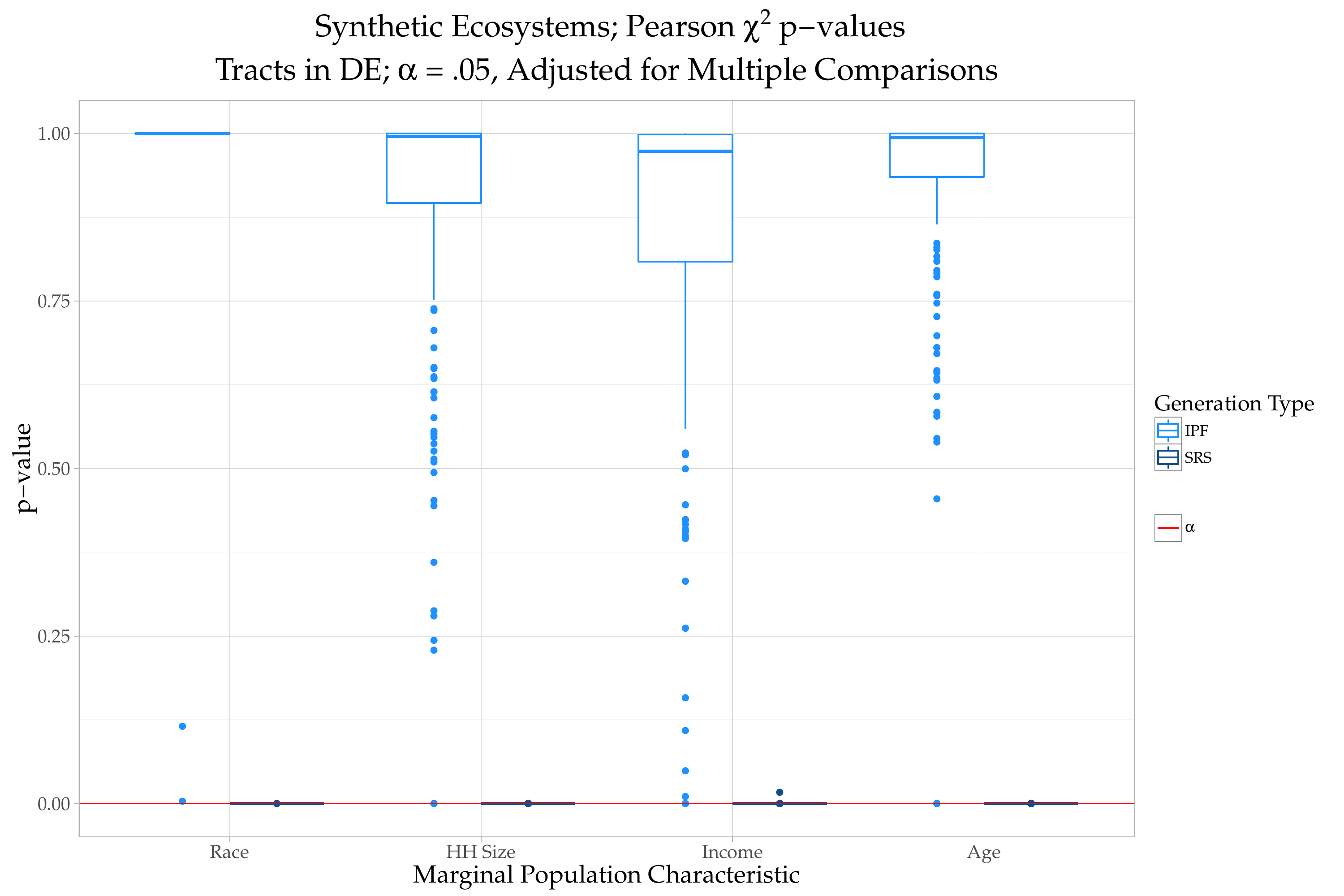}
    \caption{Boxplots of $p$-values from the $\chi^2$-test statistics, adjusted for multiple comparisons.  The test is $H_0:$  The characteristic in the synthetic ecosystem of this tract matches the true marginal totals from the input data.}\label{fig::boxplots_pval_marg}
\end{figure}

The MAE provides a summary statistic to quickly analyze the accuracy of our generated population characteristics across the different sampling methods.  Here, we compare the entire marginal distributions in order to provide a more comprehensive assessment of accuracy.  We use the Pearson $\chi^2$-statistic to evaluate whether the marginal distributions of population characteristics of our SPEW synthetic ecosystems (with SRS or IPF) have similar distributions to the true marginal distributions of age, race, income, and household size for all tracts in Delaware.  The null hypothesis of each test is that the marginal distribution of a specific synthetic agent characteristic matches the true distribution.  Because IPF for tracts within the same PUMA is initialized with the same PUMA-level joint distribution, we adjust for multiple comparisons for all tracts within a PUMA using the Bonferroni correction, repeating this adjustment for all PUMAs.  A low $p$-value indicates that it would be unlikely to observe the generated marginal distribution if the null hypothesis were true.  We note that although the Bonferroni correction is conservative, the rejected values have $p$-values so small that they be rejected with nearly any correction.  Figure \ref{fig::boxplots_pval_marg} shows the distribution of $\chi^2$-statistic $p$-values across all 214 tracts for each characteristic.

For the SRS generated characteristics, 76\% of the null hypotheses are rejected, meaning that the characteristics are distributed differently than from the input data population characteristics with 95\% confidence.  In comparison, for the IPF generated characteristics, 1.4\% of the null hypotheses are rejected (also with 95\% confidence).  This is illustrated in Figure \ref{fig::boxplots_pval_marg}.  IPF matches the true distribution of the characteristics with more accuracy than SRS.  Race, in this figure, nearly always has a $p$-value of 1 because the majority of agents in the tracts are white.  We use the $\chi^2$-statistic to flag the tracts with potential problems in our automated diagnostics.

Readers may notice that the  $p$-values in Figure \ref{fig::boxplots_pval_marg} are not uniformly distributed between 0 and 1.  This is because IPF is specifically designed to match the input marginal distributions of population characteristics.  If, instead, we were to sample population characteristics from the IPF-generated joint distributions, then we would see a uniform distribution of $p$-values.  Likewise, the SRS-generated synthetic ecosystems have population characteristics like those of the PUMS data or similar input.  In the United States, for example, tracts are subsets of PUMAs and so we would not expect a tract-level marginal distribution to match a PUMA-level distribution.  This is why all of the $p$-values for the SRS-generated tracts are so small.
  
\subsection{Assessing Spatial Sampling Approaches} \label{sec::results_spatial}
\begin{figure}[h]
  \centering
  \includegraphics[width=.9\textwidth]{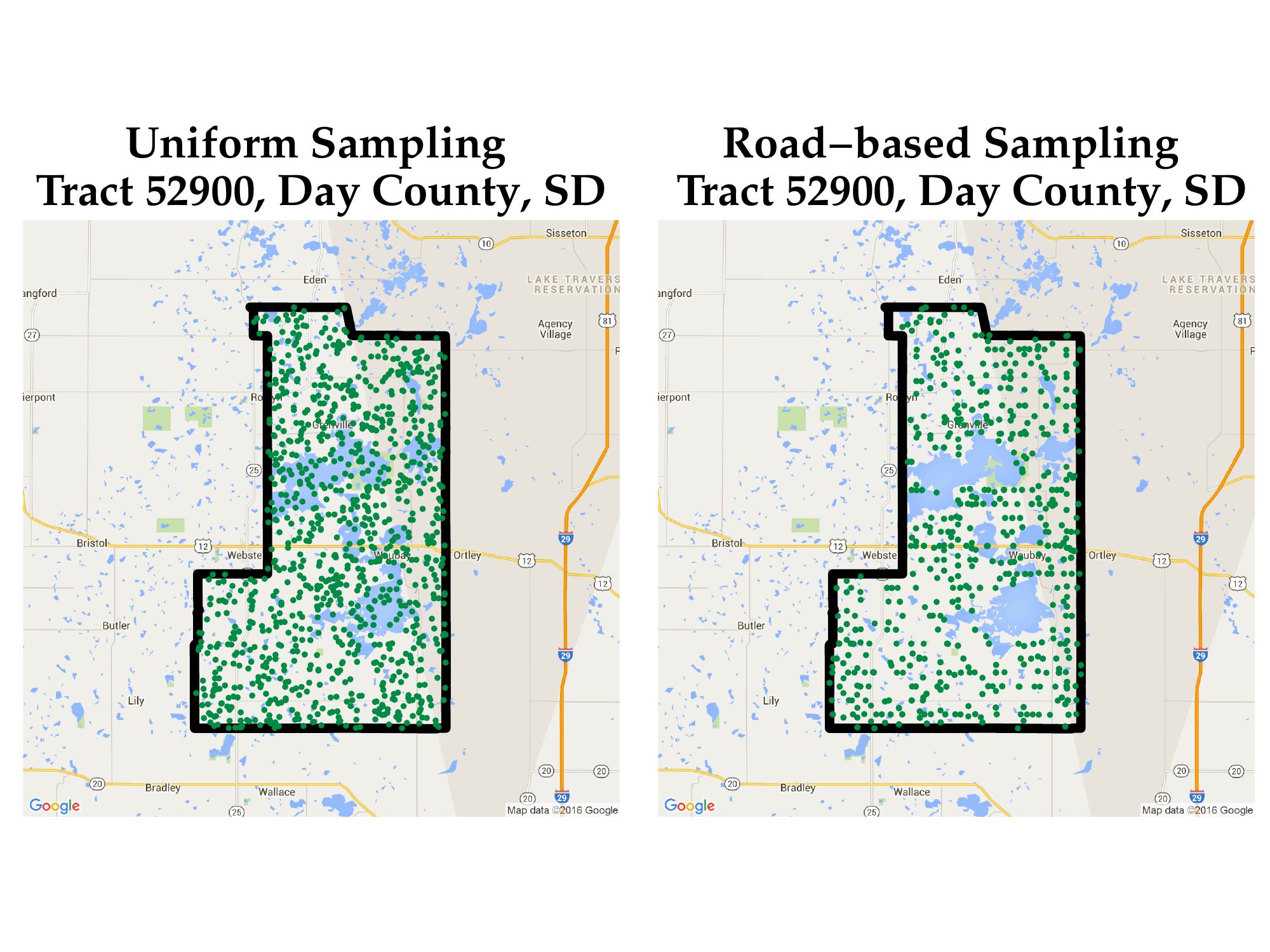}
    \caption{Comparison of uniform sampling (left) and road-based sampling (right) of households for one South Dakota census tract.  Each ($\bullet$) represents one synthetic household.  Road-based sampling avoids topographical features such as lakes while uniform sampling assigns agents to locations within lakes.}\label{fig::sd_roads}
\end{figure}

In Section \ref{sec::spatial_sampling}, we described three different methods for assigning agents to locations in SPEW.  For the reasons discussed in Section \ref{sec::spatial_sampling}, we only demonstrate the uniform sampling and road-based sampling methods in Figure \ref{fig::sd_roads}.  As shown in the figure, uniform sampling incorrectly places agents in lakes, while road-based sampling implicitly takes into account the topograhy of a geographic region, avoiding the lakes.  The maps in Figure \ref{fig::sd_roads} show this specifically for lakes, but this is true for any topographical feature that roads avoid.

Qualitatively, the road-based sampling map in Figure \ref{fig::sd_roads} appears to provide a more realistic spatial distribution of agents than the uniform sampling map.  Quantitatively, the road-based sampling should provide a better estimate of micro-level population density than uniform sampling, since the density of roads is directly related to  the density of agents.

\subsection{Computational Scalability of SPEW} \label{sec:comp-scal-spew}
The latest release of SPEW (1.2.0) synthetic ecosystems contains \spewn{} different countries and over \agentsn{} human agents.  All synthetic ecosystems were generated on the Olympus computing cluster \cite{olympus}, using the \texttt{spew} R package.  In total, the generation of all SPEW ecosystems takes a little over one hour. 

SPEW ecosystems can be divided into three different groups, based on the input data.  The first group is the United States, which is split into the 50 States, plus the District of Columbia and Puerto Rico.  For each state, we have population counts and geographies available at the tract level, so a synthetic ecosystem is generated for each tract.  In addition, we have county-level data for schools and workplaces, tract-level summary tables, and country-level road data.  With this supplementary data, we are able to use IPF for population characteristics and road-based spatial sampling.  We also included the school and workplace environmental components in the publicly available SPEW synthetic ecosystems.  

We call the second group IPUMS-I, since it is based on IPUMS-I population microdata and geographies.  IPUMS-I synthetic ecosystems are split into 73 countries, with geographical data at Administrative Level 1 (approximately the size of a state or province).  IPUMS-I synthetic ecosystems do not yet incorporate supplemental information, so the synthetic populations are generated using the default sampling methods, SRS and Uniform Sampling.  The third data group is Canada, for which we created a custom tract-level synthetic ecosystem, using data from Statistics Canada.  Information on the three groups is summarized in Table \ref{tab:spew_output}.

\begin{table}[ht]
  \centering 
  \caption{The three data groups for current SPEW synthetic ecosystems: U.S., IPUMS-I, and Canada. Count refers to how many synthetic ecosystems were generated of each type.}
  \label{tab:spew_output}
  \resizebox{\textwidth}{!}{
  \begin{tabular}{|r|r|r|r|r|r|r|}
    \hline
    Group & Count & Level & Region Level & Characteristic Sampling & Spatial Sampling & Environmental Components \\
    \hline 
    \hline
    United States & 52 & State & Census Tract & IPF & Roads & Schools, Workplaces \\ \hline
    IPUMS-I & 73 & Country & Admin. Level 1 & SRS & Uniform & None \\ \hline
    Canada & 1 & Country & Tract & SRS & Uniform & None \\ 
    \hline  
  \end{tabular}
  }
  
\end{table}

The computation of each population at the largest level (e.g states) is as follows.  Every ecosystem is split into the lowest available region level.  For example, we split California into its 8,000+ tracts.  Then, each region is generated on a single core, as in Algorithm \ref{alg::spew}. 

We adapt the parallel approach based on characteristics of the input data group.  For instance, the United States is composed of over 88,000 small tracts, so we want to use the cores of more than a single compute node.  In this case, we utilize the Message Passing Interface \citep{snir1998mpi}, implemented with the \texttt{Rmpi} package \citep{yu2002rmpi} to split the computations across multiple nodes.  In contrast, each IPUMS-I country is composed of approximately 30 regions, so generating these synthetic ecosystems only requires a single node.  In this case, we utilize the multicore functionality of the \texttt{parallel} package \citep{rcore2015}, which uses shared memory. 

Finally, we demonstrate characteristics of our computations.  In the United States, we allocate nodes based on the number of tracts in a particular state.  For instance, a large state like California is generated using 16 nodes, whereas small states like South Dakota and Delaware only use one.  In addition, since some of the regions in China and India consist of more than 100 million people, the calculations take longer than an hour for these particular regions. 

\begin{figure}
  \centering
  \label{fig::methods}
  \begin{subfigure}{.5\textwidth}
    \centering
    \includegraphics[width=\linewidth]{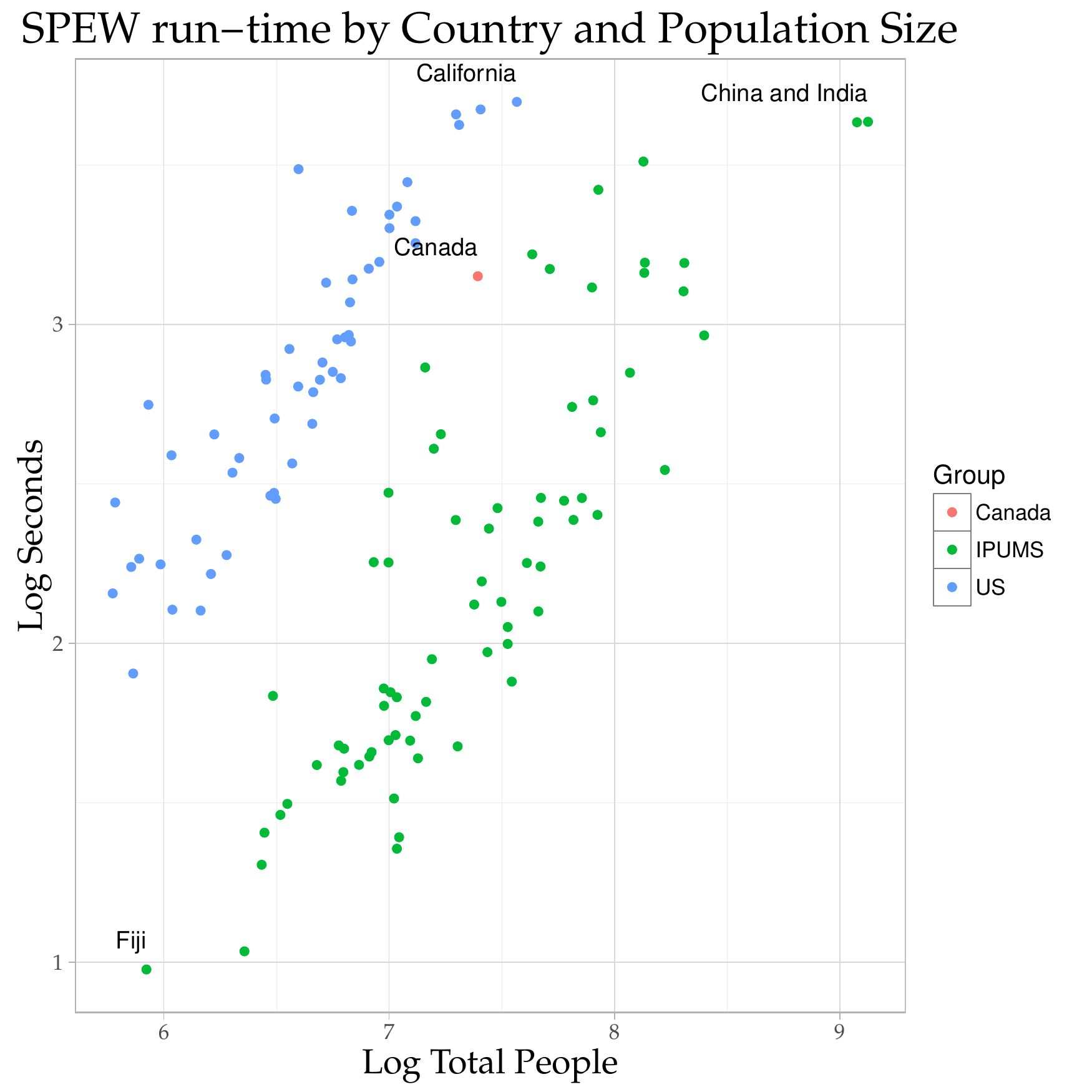}
          \caption{}  \label{fig::spew_overall}
  \end{subfigure}%
  \begin{subfigure}{.5\textwidth}
    \centering
    \includegraphics[width=\linewidth]{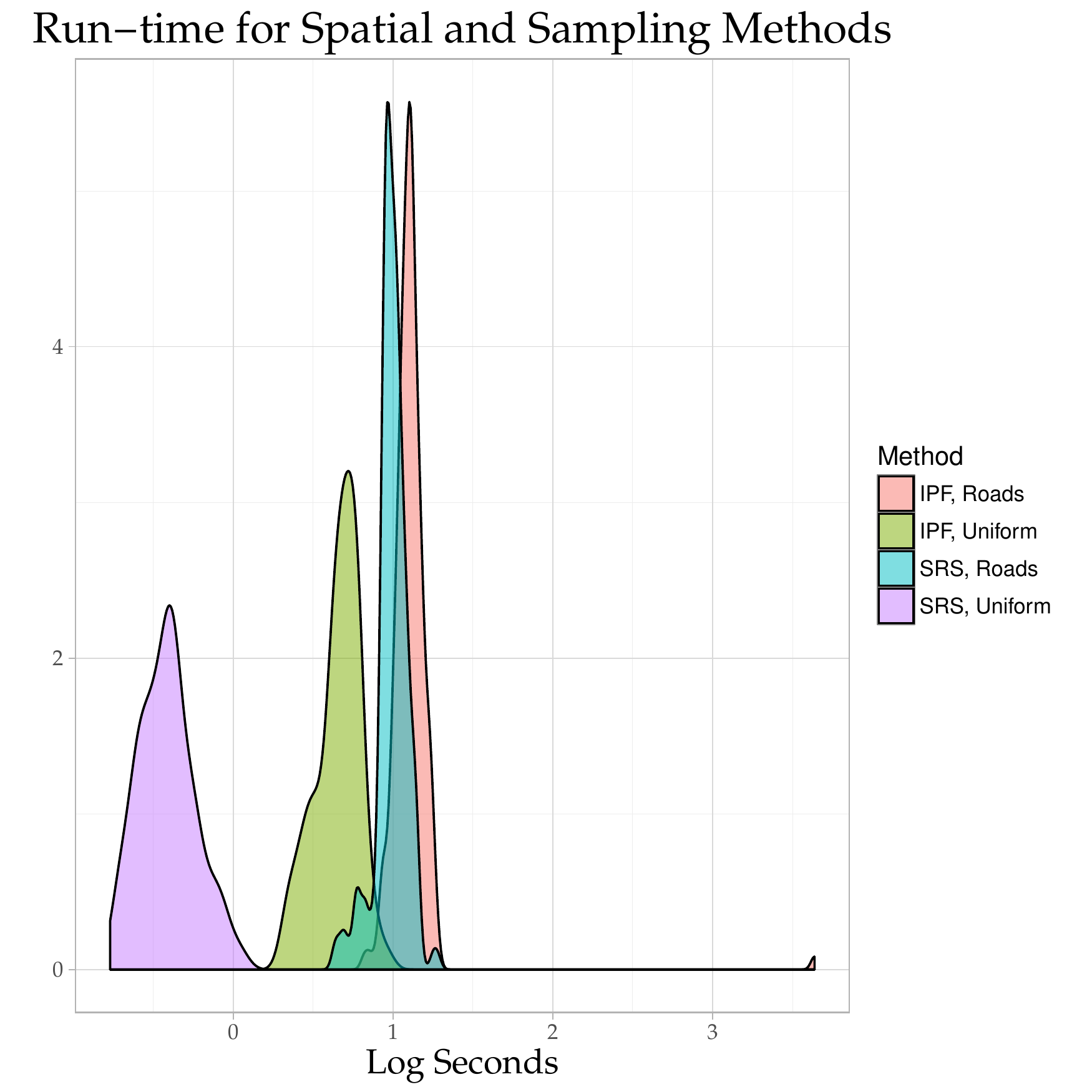}
        \caption{} \label{fig::spew_compute_methods}
  \end{subfigure}
    \caption{Left:  Log region population size vs. log run-time by input data type.  Population characteristics for both U.S. and non-U.S. regions were generated using SRS in this example.  U.S. populations were generated with road-based spatial sampling, non-U.S. sampling were generated with uniform spatial sampling, and Canada was generated from input data from Statistics Canada.  Additionally, agents were assigned to schools and workplaces for U.S. regions, while agents were not assigned to any environmental components for non-U.S. regions.  Right:  Density estimates of SPEW log run-time by population characteristic and spatial sampling methods for census tracts in Delaware.  IPF appears to have lower variance in log run-time than SRS methods, adjusting for the spatial sampling method, but takes longer on average than SRS.  As expected, road-based sampling takes longer to run on average than uniform sampling of agent locations.} 
\end{figure}

Figure \ref{fig::spew_overall} shows the relationship between region size (number of people) and run-time, by input data type.  The overall run-time is clearly a function of total population size.  Generally, a single U.S. synthetic ecosystem takes much longer than the non-U.S. ecosystems.  This is because the U.S. ecosystems are more complex, containing an assignment of agents to schools and workplaces and road-based (rather than uniform) spatial sampling. 

In Figure \ref{fig::spew_compute_methods}, we compare the run-time of SPEW for tracts in Delaware, conditional on agent characteristic sampling type (left) and spatial sampling method (right). Road-based spatial sampling substantially increases the computational cost of generating synthetic ecosystems, largely due to reading and manipulating large shapefiles.  IPF is more computationally intensive than SRS but has a smaller variance in log run-time. 

\begin{figure}
  \centering
\begin{subfigure}{.5\textwidth}
  \centering
  \includegraphics[width=.9\linewidth]{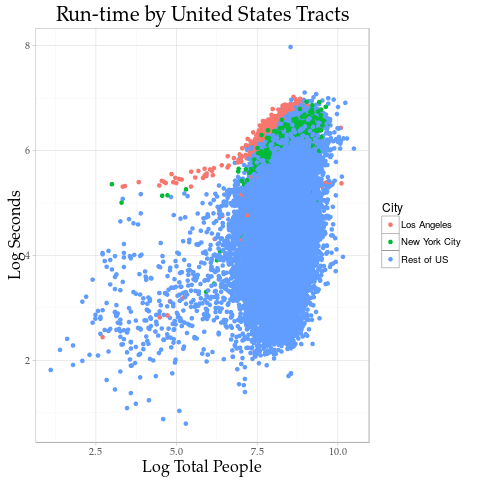}
      \caption{} \label{fig::spew_region_runtime}
\end{subfigure}%
\begin{subfigure}{.5\textwidth}
  \centering
  \includegraphics[width=.9\linewidth]{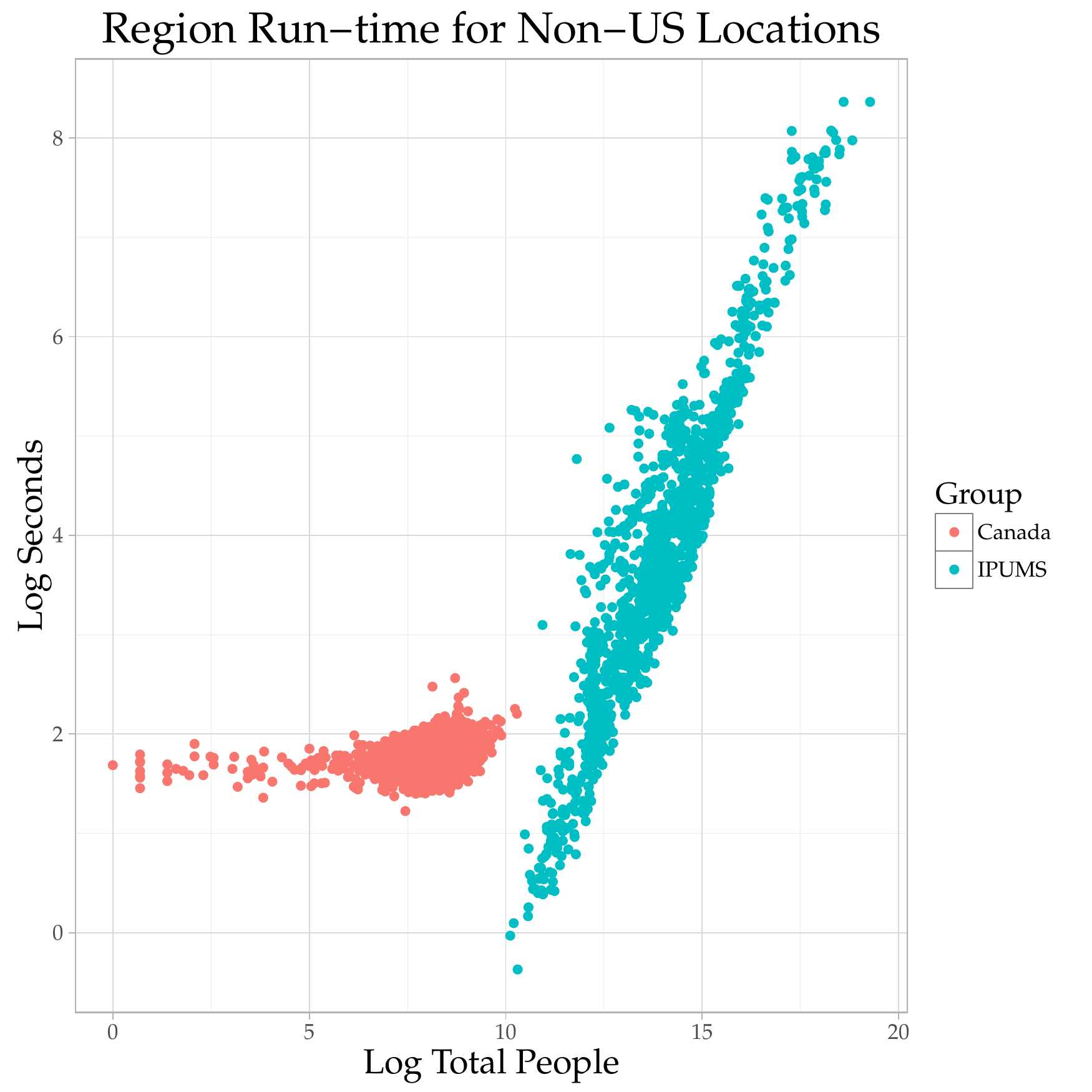}
    \caption{}\label{fig::spew_region_nonus}
\end{subfigure}
  \caption{Left:  Total number of agents in U.S. generated region vs. run-time in seconds.  Highlighted are tracts Los Angeles, CA and New York, NY, each of which has an unusually large number of roads in their surrounding county.  Regions with a large number of roads take longer to generate due to the size of the input roads data.  Right:  Total number of agents in non-U.S. generated regions vs. run-time in seconds for regions in Canada vs. IPUMS-I regions.  There is a linear relationship between log total people and log run-time for IPUMS-I regions.  Canadian regions typically have a smaller number of people but there is no obvious relationship between log total people and log run-time.}
\end{figure}

The variance in run-time with road-based sampling is large, likely because of the large variance of the number of roads within a given tract, directly effecting run-time.  Figure \ref{fig::spew_region_runtime} demonstrates this by highlighting tracts with a large number of roads.  Specifically, we have highlighted the tracts in Los Angeles and New York City because there are more roads here than in a typical tract.  Run-times in Los Angeles and New York are much larger, even for tracts with equal sized populations.  As such, when allocating computational resources for U.S. regions, the number of roads in each tract is an important variable. 

In contrast, Figure \ref{fig::spew_region_nonus} shows that run-time of non-U.S. regions is only affected by population size.  With the exception of Canada, the log run-time for IPUMS-I grows linearly with the total population size.  

\section{Discussion}\label{sec::discussion}

We extend previous approaches for generating synthetic ecosystems for use in ABMs.  We introduce a framework for generating synthetic ecosystems called SPEW.  SPEW processes essential input (population counts, geography, and population characteristics) and outputs a synthetic ecosystem containing agents, their locations, their characteristics, and additional information about their environment.  Within the SPEW framework, we may use different approaches for sampling population characteristics and sampling from a spatial distribution.  In this paper, we assess different methods for sampling population characteristics, statistically comparing the distributions of population characteristics in our synthetic ecosystems to the truth.  We also examine the benefits of using more sophisticated spatial sampling of agent locations.  Next, we extend methods for assigning agents to environmental components.  Finally, we study the computational tractability of generating numerous synthetic ecosystems from different data sources and various geographic regions.  

We discuss three approaches to sampling population characteristics:  SRS, MM,  and IPF.  We provide a detailed comparison of SRS and IPF for geographic regions in the United States.  In particular, we compare the distribution of population characteristics from the synthetic ecosystems generated from SRS and IPF to those of the true marginal distributions from the U.S. Census Bureau.   We compute the associated $\chi^2$-statistics and $p$-values and find that approximately 99\% of IPF-generated tracts have  population characteristics that match the truth, whereas only 25\% of the SRS-generated tracts match the truth.  As a result, we favor using IPF to generate population characteristics, given the appropriate input data sources are available.

We examine two methods for sampling locations of agents:  uniform and road-based sampling.  With uniform sampling, agents are assigned to random locations within the boundaries of the regions.  For road-based sampling, agents are assigned locations near to roads.  We find that while uniform sampling maintains macro-level population densities, it does not provide a  realistic micro-level population density of agents within regions.  Additionally, this method is subject to obvious errors, such as placing agents in lakes.  We advocate for using a known spatial distribution of agent locations such as sampling from non-interstate roads.  

For ABMs, it is important for agents to be assigned to components in their environment in order to simulate their interactions.  We provide a flexible function to assign agents to environmental components such as schools, workplaces, and places of worship.  This function is based on both distance of agent to environmental component and the capacity of the environmental component.  This capability is illustrated when we assign people in a neighborhood to places of worship with different capacities in Figure \ref{fig::churches}.

Generating synthetic ecosystems for the world is inherently a computationally intensive task, heavily dependent on the type of input data, the  sampling approach for population characteristics,  the sampling approach for agent locations, and the number of agents per region.  First, we find the run-time for regions in the United States is typically larger than non-U.S. countries, after adjusting for population size, due to the availability of more detailed input data and hence more sophisticated sampling methods.  Second, we find that the run-time for IPF-generated synthetic ecosystems is higher on average than SRS-generated synthetic ecosystems, but the variance of run-time for IPF is less than that of SRS.  Additionally, we find that while road-based sampling provides a more realistic spatial distribution of agents, it is the most computationally intensive of any of our current sampling methods, especially when there is a high density of roads in a region such as large cities in the U.S., like New York or Los Angeles.  Finally, and perhaps most clearly, there is a positive linear relationship between the log number of agents and log run-time.

The SPEW framework may be used to generate any agent type given the appropriate input data for any geographic region.  We implement the SPEW framework in our \texttt{R} package.  To date, we have used SPEW to produce \agentsn{} human agents across \spewn{} countries.  These completed synthetic ecosystems are available publicly for agent-based modellers and other researchers at \url{http://data.olympus.psc.edu/syneco/spew_1.1.0/}.

Because of the flexibility of the SPEW framework, we may generate custom synthetic ecosystems for researchers with proprietary input data or specific modeling requests with regards to sampling methods.  For example, we have generated a custom synthetic ecosystem for California that specifically emphasized the accuracy of the Hispanic population using alternative input data from the state of California.  Additionally, our Canadian synthetic ecosystem used data from Statistics Canada, which did not have publicly available microdata.   As a final example, we have generated a custom synthetic ecosystem for a state in India using proprietary data for a collaborator interested in modeling the spread of AIDS.

Our SPEW synthetic ecosystems have broad impact for researchers who use ABMs.  First, SPEW provides a standardized output that may be used to generate synthetic ecosystems across countries, years, and types of agents.  For instance, even though the input data for the United States is substantially different than that of India, the resulting synthetic ecosystems are identical in format.  The years are important as researchers may want to reproduce a synthetic ecosystem for an event that took place in the past (e.g. the 2009 influenza pandemic).  The SPEW framework may be used to generate new synthetic ecosystems in the future when new data becomes available.  As we have seen with recent epidemics such as Ebola, Dengue, and Zika, diseases are transmitted not only through humans, but through disease vectors such as mosquitoes.  Given the appropriate input data, the SPEW framework allows us to generate synthetic ecosystems with these non-human agents.

Moreover, SPEW produces high-resolution synthetic ecosystems given high-resolution input data.  As such, researchers will have access to higher quality synthetic ecosystems for use as input in their ABMs given the appropriate input data sources.  Additionally, researchers with proprietary data sources (e.g. ESRI) may incorporate this information into their synthetic ecosystems via SPEW's modular framework.  More generally, environmental components may be added via SPEW to allow a richer set of interactions within ABMs.  Finally, SPEW provides a standard for synthetic ecosystem generation.

Moving forward, we want to automatically determine the optimal computational approach within SPEW given the machine constraints and input data.  Specifically, we want to automatically determine the optimal parallelization approach based on characteristics of the synthetic ecosystem to be generated.  For example, we might want to know the optimal method for splitting independent regions accross compute cores.  To determine this, we must understand the expected run-time for each region and take into account the processing and data storage capabilities of the user's computing system.

In future work, we plan to publicly release SPEW synthetic ecosystems for disease vectors such as mosquitoes, livestock, or poultry, in addition to the already released human synthetic ecosystems.  We plan to explore new sampling methods to include completely synthetic data rather than sampling from existing microdata.  Additionally, we hope to identify data sources that allow for more accurate spatial sampling outside of the United States (e.g. Open Street Maps for road-based sampling of non-U.S. countries).  As new versions of existing data sources (and entirely new data sources) become available, we will continue to release updated SPEW synthetic ecosystems incorporating these new sources of data.  Moreover, we hope to include more environmental components where agents' activities may occur such as airports and restaurants, as these loci of transmission are essential in disease modeling.   Finally, we note that in some regions, the essential input data may never become available.  Since these regions are often third-world countries and thus may be most susceptible to new disease outbreaks, it is important that we find ways to generate synthetic ecosystems for these regions.  Doing so will allow us to generate a high-resolution synthetic ecosystem of the entire world.

\bibliographystyle{apa}
\bibliography{masterBib}

\appendix

\section{}\label{sec::datalist}
\bigskip
\begin{center}
{\large\bf SUPPLEMENTARY MATERIAL}
\end{center}

\begin{enumerate}
\item \textbf{Synthetic Populations and Ecosystems of the World}.  All ecosystems from SPEW are available at \url{http://data.olympus.psc.edu/syneco/spew_1.1.0/}  (\texttt{.zip, .csv}).

\item \textbf{Code}. The \texttt{R} package is available online at: \url{https://github.com/leerichardson/spew}.

\item \textbf{Input Data} to SPEW.
  \begin{enumerate}
  \item Counts.  These are population totals for country administrative regions.  Source include the U.S. Census ACS Summary Files and GeoHive. (\texttt{.csv})
    
  \item Geographies.  These are the digital boundaries for country regions. Sources include the U.S. Census TIGER and IPUMS-I microdata. (\texttt{.shp, .shx, .dbf, .prj}).  
    
  \item Microdata.  These are the individual-level data for a given region.  Sources include the U.S. Census ACS PUMS and IPUMS-I shapefiles. (\texttt{.csv})
    
  \item Supplementary Data.  These include regional data such as schools, workplaces, or places of worship.  Sources include NCES, Esri, and OpenStreetMaps. (\texttt{.csv})
  \end{enumerate}

\end{enumerate}

\section{Statistical Details}\label{app::ipf}

\subsection{Iterative Proportional Fitting}
  The two-step method of \cite{Beckman1996} works as follows:
  \begin{enumerate}
    \item Estimate the joint distribution of population characteristics using IPF and a set of marginal totals.
    \item Sample records/agents from the microdata using probabilities from the joint distribution as weights.
  \end{enumerate}

  Following \cite{Beckman1996}, our notation is:

  \begin{itemize}
    \item $n: $ total number of observations in the table
    \item $m: $ number of  demographic characteristics (dimensions of the contingency table)
    \item $n_j: $ number of categories for the $j^{th}$ demographic. $j = 1, 2, ..., m$
    \item $i_j: $ value of the $j^{th}$ demographic. $i_j = 1, 2, ..., n_j$ 
    \item $p_{i_1, i_2, ..., i_m} = \frac{n_{i_1, i_2, ..., i_m}}{n}: $ proportion of observations in an individual cell
    \item $T_k^{(j)}: $ marginal totals for $k^{th}$ category of $j^{th}$ demographic. $k = 1, 2, ..., n_j$
  \end{itemize}

  So for all $j$, we have:

  \begin{equation}
    n = \sum_{k = 1}^{n_j} T_k^{(j)}
  \end{equation}

  IPF updates the contingency table until the marginal totals are within a tolerance of the known marginals. Each update is called an iteration. Let $p_{i_1, i_2, ..., i_m}^{(t)}$ denote the estimation of the cell $(i_1, i_2, ..., i_m)$ during iteration $t$.

  The initial contingency table for IPF is:

  \begin{equation*}
    p_{i_1, 1_2, ...i_m}^{(0)} = p_{i_1, i_2,..., i_m}.
  \end{equation*}

  This is the seed contingency table and incorporates prior information into the final contingency table.  In practice, the seed table originates from the microdata.   For instance if the microdata has four males heading three-person households, aged 30-34, and earning \$100,000 dollars a year, then:

  \begin{align*}
   p_{i_{gender}, i_{hhsize}, i_{age}, i_{income}} &= 4.
  \end{align*}
  Each iteration goes through each margin, and updates the estimated proportion $\hat{p}_{i_1, i_2, ..., i_m}$. Specifically, for each of the $j$ margins, we update the $k^{th}$ category by:

  \begin{equation*}
    p_{i_1, i_2, ...,i_j = k, ...,i_m}^{(t)} = p_{i_1, i_2, ...,i_j = k, ...,i_m}^{(t - 1)} \frac{T_{k}^{(j)} / n}{\sum_{i = 1}^{n_1} \sum_{i = 1}^{n_2} ... \sum_{i = 1}^{n_m} p_{i_1, i_2, ...,i_j = k, ...,i_m}^{(t - 1)}}.
  \end{equation*}

  We continue iterations until the tolerance is reached.  \cite{Beckman1996} reports that the procedure typically converges in 10-20 iterations.

  Step 2 samples households in proportion to the probabilities in the contingency table. The probabilties determine how many households of each demographic combination should be sampled. For each demographic combination, probabilities are assigned to each microdata household, based on how ``close'' the household is, in terms of demographics. The closeness of each household is determined by the distance function:

  \begin{equation*}
    D(p, c) = w_p \prod_{i \in J} (1 - |\frac{d_i^p - d_i^c}{r_i}|^k) \times \prod_{i \notin J} (1 - (\delta(d_i^p, d_i^c)).
  \end{equation*}

  With the following notation:

  \begin{itemize}
    \item $p: $ household from microdata
    \item $c: $ cell from contintency table
    \item $J: $ set of ordinal and continuous variables. $J^C$ is the set of categorical variables.
    \item $d_i^p: $ value of $i^{th}$ demographic for household $p$
    \item $d_i^c: $ value of $i^{th}$ demographic of cell type $c$
    \item $r_i: $ range of demographic $i$ in the PUMS
      \item $k$ a weight for the current demographic category
    \item $w_p: $ weight from household $p$
    \item $\delta(d_i^p, d_i^c) = \begin{cases} \alpha & d_i^p = d_i^c \\ 1 - \alpha & d_i^p \neq d_i^c \end{cases} $
  \end{itemize}

  Note that when $\alpha = 0$ and $k \to 0$, $D(p, c)$ is a 0-1 loss function, which means we are only sampling from records that eactly match those in the cell of the contingency table from IPF. 

  Finally, each record is sampled according to the following weights:

  \begin{equation*}
    \mathbb{P}(\text{Select Household } p) = \frac{D(p, c)}{\sum_j D(j, c)}.
  \end{equation*}
 
  \subsection{Moment Matching}

  Moment matching is a method to assign weights to records in the microdata so that after sampling, the moment of the resulting population match the moment from prior knowledge.  Assume we have access to the first moment of a population characteristic in region $r$, but the distribution of that characteristic is unknown. Denote this moment as $M_r$. Moreover, we assume that there exists a microdata, with $m$ distinct characteristic values $\textbf{n}=(n_1, \dots, n_m)^T$ where each $n_i > 0$. Then after, sampling, we expect that $M_r$ is still the first moment of our resulting synthetic ecosystem.  Denote weights of the microdata by $\textbf{w}=(w_1, \dots, w_N)^T. $ Let $w_i \ge 0$ for $i= 1, \dots, N$, $\sum_{i=1}^N w_i =1$, and $\sum_{i=1}^N w_in_i=M_r$ denote the constraints.  This formulation alone has infinitely many solutions.  To settle on a particular value, we form a quadratic program and minimize  $$f(w) =\frac{1}{2} ||\textbf{w}||_2^2,$$ where $||\cdot||_2$ is the $L^2$ norm. We note that one reason for this proposed method is due to its is simplicity.  One advantage to using the $L_2$ norm is that we use many of the possible microdata records instead of just using the same ones over and over.

We use this set up to form a quadratic program.  Our objective function $f$ and the constraints are as follows

\begin{align}
f(\textbf{w}) &= \frac{1}{2}||\textbf{w}||^2  \label{obj} \\
{\rm subject\;\; to }\; & - x_i \le 0 \text{ for } i=1, \dots, N ;\label{ineq} \\
& \sum_{i=1}^N w_i -1 = 0; \label{l1} \\
&\sum_{i=1}^N n_i w_i - M_r. \label{l2}=0.
\end{align}

Equations \ref{l1} and \ref{l2} imply that 

\begin{align}\label{l3}
\textbf{Aw} = \textbf{b}\text{, where}
\end{align}
\begin{align*}
\textbf{A} = \left [ 
\begin{array}{ccc}
  1 & \dots & 1 \\
  n_1 & \dots & n_N 
\end{array}
 \right ]\text{, } \textbf{w}=
\left [
\begin{array}{c}
  w_1 \\
  \vdots \\
  w_n
\end{array}
\right ] \text{, } \textbf{b} =
\left [
\begin{array}{c}
  1 \\
  M_R
\end{array}
\right ] .
\end{align*}

It is clear from Equations \ref{obj}-\ref{l3}, that we have a quadratic program, which may be written as 
\begin{align}
\min_{\mathbf{w} \in \mathbb{R}^N} \frac{1}{2} \mathbf{w}^T \mathbf{w}, \textnormal{ with } \mathbf{A} \mathbf{w} = \mathbf{b}, \mathbf{w} \ge 0.
\label{qp}
\end{align}

To solve the quadratic program in \ref{qp}, we utilize the \texttt{quadprog} package in \texttt{R}, which implements the algorithm of Goldfarband Idnani related by \cite{powell_1985}.

\end{document}